\documentclass[journal]{IEEEtran}
\usepackage{amsmath,amsfonts}
\usepackage{array}
\usepackage[caption=false,font=normalsize,labelfont=sf,textfont=sf]{subfig}
\usepackage{textcomp}
\usepackage{stfloats}
\usepackage{url}
\usepackage{verbatim}
\usepackage{graphicx}
\usepackage{cite}
\usepackage{xcolor}
\usepackage{amssymb}
\usepackage{bm}
\usepackage{algpseudocode}
\usepackage{makecell}
\usepackage{enumerate}
\usepackage{multirow}
\usepackage{diagbox}
\usepackage[ruled,linesnumbered]{algorithm2e}
\usepackage{amsmath,bm,graphicx,multirow,bm,bbm,amssymb}
\begin{document}

\title{Joint User Scheduling, Power Allocation, and Rate Control for MC-RSMA in URLLC Services}

\author{Xiaoyu~Ou,~\IEEEmembership{Student~Member,~IEEE,}
    Shuping~Dang,~\IEEEmembership{Senior~Member,~IEEE,}
    Zhihan~Ren,~\IEEEmembership{Student~Member,~IEEE,}
 	and~Angela~Doufexi,~\IEEEmembership{Member,~IEEE,}
 	
\thanks{This work was supported by the University of Bristol Scholarship for the Electrical and Electronic Engineering PhD.}
\thanks{Xiaoyu Ou, Shuping Dang, Zhihan, Ren, and Angela Doufexi are with the School of Electrical, Electronic and Mechanical Engineering, University of Bristol, Bristol BS8 1UB, U.K. (e-mail: xiaoyu.ou@bristol.ac.uk; shuping.dang@bristol.ac.uk; z.ren@bristol.ac.uk; a.doufexi@bristol.ac.uk).}}



\maketitle

\begin{abstract}
This paper investigates the resource management problem in multi-carrier rate-splitting multiple access (MC-RSMA) systems with imperfect channel state information (CSI) and successive interference cancellation (SIC) for ultra-reliable and low-latency communications (URLLC) applications. To explore the trade-off between the decoding error probability and achievable rate, effective throughput (ET) is adopted as the utility function in this study. Then, a mixed-integer non-convex problem is formulated, where power allocation, rate control, and user grouping are jointly taken into consideration. To tackle this problem, we approximate the achievable ET using a lower bound and then develop a decomposition method to decouple optimization variables. Specifically, for a given user grouping scheme, an iteration-based concave-convex programming (CCCP) method and an iteration-free lower-bound approximation (LBA) method are proposed for power allocation and rate control. Next, a greedy search-based scheme and a heuristic grouping scheme are developed for the user-grouping problem. The simulation results verify the effectiveness of the CCCP and LBA methods in power allocation and rate control and the greedy search-based and heuristic grouping methods in user grouping. Besides, the superiority of RSMA for URLLC services is demonstrated when compared to spatial division multiple access.
\end{abstract}


\section{Introduction}
\IEEEPARstart{N}{on-orthogonal} multiple access (NOMA) has been emerging as a promising technology for fifth-generation (5G) and 6G networks \cite{ref1}, \cite{ref2}. NOMA enables higher network throughput, energy efficiency, and spectral efficiency by transmitting multiple signals on a single resource block (RB). Meanwhile, the reduced energy consumption of NOMA makes it a promising candidate for realizing green and sustainable communication networks. In addition, these advantages have been proven effective in mission-critical URLLC applications in improving reliability and reducing latency \cite{ref3}. However, the superposition coding and successive interference cancellation (SIC)-based NOMA technology impose high computational requirements on the receiver \cite{ref4}. Meanwhile, multi-layer SIC operations of NOMA could severely degrade the reliability of the system, especially when it interacts with URLLC services, whose decoding error probability matters due to the finite blocklength (FBL) \cite{ref5}.

To this end, a new extension of NOMA termed rate-splitting multiple access (RSMA) was recently proposed to overcome the inherent disadvantages of NOMA \cite{ref6}. In downlink RSMA, messages aimed to be received by multiple users are split into common parts and private parts. The common parts are combined and encoded into a common stream that is mainly utilized for interference management, while the private parts are independently encoded into private streams. In this way, only one-layer SIC operation is required in the decoding phase after users decode the common stream, which effectively reduces the decoding complexity at the receiver. Besides, the flexible interference management improves the spectral efficiency, throughput, and latency performance of RSMA compared to spatial division multiple access (SDMA) in low-interference scenarios and NOMA in high-interference scenarios \cite{ref7, ref8, ref9, ref10}. Therefore, under the same quality of service (QoS) constraints, RSMA has great potential to reduce power consumption and computational resources by virtue of its low-complexity system architecture. 

\subsection{Related Works}
Due to the impact of FBL, the traditional Shannon capacity needs to be replaced by a new approximate metric proposed in \cite{ref11} for short packet communications (SPC)-driven URLLC applications:
\begin{equation}\label{eq1}
	R \approx \log_2(1 + \gamma) - \frac{Q^{-1}(\varepsilon^{\mathrm{th}})}{\ln2}\sqrt{\frac{V}{N}},
\end{equation}
where $\gamma$ represents the received signal-to-interference-plus-noise-ratio (SINR); $V = 1 - (1 + \gamma)^2$ characterizing channel dispersion; $N$ is the channel blocklength, $Q(x) = \int_{x}^{\infty}\frac{1}{\sqrt{2\pi}} e^{-\frac{t^2}{2}} dt$ refers to the Gauss function, and $Q^{-1}(x)$ stands for its inverse; $\varepsilon^{\mathrm{th}}$ stands for the decoding error threshold.

Based on the approximation given in (\ref{eq1}), the problem of physical layer resource management for URLLC services has received extensive academic attention. Specifically, the work reported in \cite{ref12} studied the maximization of weighted throughput and power minimization for secure IoT communication systems with URLLC requirements, while the work in\cite{ref13} focused on beamforming design for downlink multi-user URLLC systems. Besides, a robust beamforming design scheme for URLLC with imperfect channel state information (CSI) at the transmitter (CSIT) was proposed in \cite{ref14}. In \cite{ref15}, the authors investigated conjugate beamforming for cell-free massive multiple-input multiple-output (MIMO) URLLC systems. The authors in \cite{ref16} investigated resource allocation problems for unmanned aerial vehicle (UAV)-assisted uplink URLLC transmission, where UAV's position, height, and beam patterns are optimized to minimize the overall uplink transmit power. Then, the authors extend their works in \cite{ref16} to reconfigurable intelligent surface (RIS)-assist URLLC systems aided by UAV relays in \cite{ref17}. 

Motivated by the performance superiority of NOMA, the works in \cite{ref18,ref19,ref20,ref21,ref22,ref23,ref24,ref25,ref26} have explored its potential applications for URLLC. In \cite{ref18}, the authors introduced NOMA into an SPC scenario for the first time, where rate, power, and channel blocklength, were jointly optimized to maximize the effective throughput (ET). The authors in \cite{ref19} considered heterogeneous receiver configurations under different interference mitigation schemes for a transmission energy minimization problem in two-user downlink NOMA systems with heterogeneous latency constraints. Meanwhile, they investigated offline packet scheduling for downlink hybrid NOMA systems with two heterogeneous latency-constrained users in \cite{ref20}. The work in \cite{ref21} studied the decoding error probability and power allocation optimization problem of NOMA with FBL transmission, which aims at maximizing the ET of the central user under the minimum-required ET constraint of cell-edge users. The authors in \cite{ref22} analyzed the SPC performance of MIMO-NOMA systems using the average block error rate (BLER) and minimum blocklength as performance indicators. The cooperative NOMA (C-NOMA) was considered in \cite{ref23} for SPC, where user grouping and resource allocation are optimized to maximize fair throughput. Additionally, \cite{ref24} explored the cognitive radio (CR)-inspired NOMA in SPC, while co-existence problems between URLLC and eMBB services in NOMA systems were investigated in \cite{ref25} and \cite{ref26}.

Additionally, RSMA has drawn much attention from researchers in recent years. Research in \cite{ref27} studied the weighted sum-rate (WSR) maximization problems of downlink single-input single-output (SISO) and multiple input and single output (MISO) RSMA systems, while \cite{ref28} focused on MIMO RSMA systems. The simultaneous wireless information and power transfer (SWIPT)-assisted MISO RSMA system was proposed in \cite{ref29} to minimize power consumption. The authors in \cite{ref30} investigated the max-min fairness (MMF) problem in two-user RSMA-based coordinated direct and relay transmission systems. Besides, the impact of imperfect CSI on RSMA systems was studied in \cite{ref31,ref32,ref33}. The work in \cite{ref31} investigated the MMF precoding problem in RSMA systems with imperfect CSIT and imperfect CSI at the receiver (CSIR). The work in \cite{ref32} studied the WSR maximization problem for RSMA in downlink multi-user (MU) systems with both perfect and imperfect CSIT. Meanwhile, the work in \cite{ref33} investigated the effective secrecy throughput of uplink UAV-assist RSMA systems with imperfect CSI.

Recently, RSMA has been introduced into FBL-enabled SPC scenarios in \cite{ref34,ref35,ref36,ref37,ref38,ref39,ref40}. Specifically, the sum rate and MMF optimization problems were respectively studied in \cite{ref34} and \cite{ref35} for downlink MISO RSMA systems with FBL constraints. Then, the work in \cite{ref36} investigated ET maximization problems for single-carrier (SC) and multi-carrier (MC) MISO RSMA-URLLC systems, respectively. Next, ET maximization problems for downlink two-group RSMA systems and uplink two-user RSMA systems with FBL constraints were studied in \cite{ref37} and \cite{ref38}, respectively. Next, the applications of RIS-assist downlink RSMA in URLLC were studied in works \cite{ref39} and \cite{ref40}. In detail, the authors in \cite{ref39} proposed an optimization framework that can provide suboptimal solutions for RIS-assist RSMA-URLLC systems, in which the objective and/or constraints are linear functions of the rates and/or energy efficiency of users. The authors in \cite{ref40} studied the spectral-efficient resource allocation for RIS-assist RSMA-URLLC systems, where precoder design, blocklength, and passive beamforming of RIS were optimized to maximize the sum rate.

\subsection{Motivations}
The interaction between RSMA and SPC for URLLC has been extensively studied in \cite{ref34,ref35,ref36,ref37,ref38,ref39,ref40}. However, further investigations are necessary to address existing deficiencies. For example, first, the time-varying channels pose challenges to CSI acquisition, especially in multi-antenna systems. Meanwhile, considering the limited pilot sequences for channel estimation, achieving perfect CSI is unrealistic in URLLC. Therefore, imperfect CSI needs to be taken into account. This means that the perfect SIC cannot be guaranteed due to channel errors. Thus, residual interference after SIC operations will further degrade system performance. However, these conditions have not been studied in the aforementioned works. Second, the optimization methods employed in these studies are based on successive convex approximation (SCA) \cite{ref34, ref35, ref36, ref37, ref38, ref39, ref40}, or alternative optimization (AO) \cite{ref37, ref38, ref39, ref40}. These methods require at least one-layer iteration to obtain suboptimal solutions, resulting in high computational complexity that would affect the latency of URLLC. Third, in NOMA/RSMA, inter-user interference (IUI) increases significantly with the number of multiplexed users, which severely degrades the efficiency and reliability of systems. Therefore, the way of integrating orthogonal multiple access, i.e., MC systems, has great potential to enable massive access in NOMA/RSMA. To this end, user grouping is an additional problem that needs to be addressed. However, previous studies reported in \cite{ref37} and \cite{ref38} only considered MC-RSMA transmission but did not provide any insight into user grouping in MC-RSMA systems.

\subsection{Contributions}
Motivated by the above deficiencies, the current work is devoted to optimize rate control, power allocation, and user grouping to maximize the sum ET of downlink MC-RSMA-URLLC systems. Meanwhile, considering the imperfect channel estimation, a more practical transmission scenario with imperfect CSI and SIC is considered. The major contributions of this work are summarized as follows:
\begin{itemize}
\item We utilize ET as the performance metric in downlink MC-RSMA systems with imperfect CSI and SIC and formulate a mixed-integer non-convex problem. For simplicity, we first approximate the achievable ET by using a lower bound and utilize a decomposition method to tackle the user grouping problem and the joint power allocation and rate control problem separately.
\item For a given user grouping scheme, we transform the problem into a difference of convex (DC) form and use concave-convex programming (CCCP) to obtain local optima. In addition, we further propose an iteration-free lower bound approximation (LBA) method to reduce complexity and processing delay.
\item For the user grouping problem, we first propose a greedy search scheme as a performance baseline. Then, we develop a heuristic grouping scheme to obtain unique insight into the user grouping problem for MC-RSMA systems. The basic principle is to associate users with lower channel correlation coefficients in the same group.
\item As verified by simulations, the proposed CCCP and LBA methods show near-optimal performance in rate control and power allocation. Meanwhile, the proposed greedy search and heuristic search methods show higher performance compared to the random method. In addition, the superiority of RSMA for URLLC services is also verified compared to SDMA.
\end{itemize}

The remainder of this paper is organized as follows: Section~\ref{Section II} presents the system model and formulates the optimization problem. Section~\ref{Section III} corresponds to details of the proposed decomposition method. Section~\ref{Section V} and Section~\ref{Section VI} provide the simulation results and conclusions, respectively.

\section{System Model and Problem Formulation}\label{Section II}
\subsection{System Framework}

This work considers a downlink URLLC scenario with MC-RSMA transmission, where a base station (BS) equipped with $M_t$ antennas serves $K$ single-antenna users ($M_t > K$). Assume that the total bandwidth of the system, denoted by $B$, is equally split into $J$ subcarriers; thus, the available bandwidth of each subcarrier is $B_j = B/J$. Then, the $K$ users need to be scheduled into $J$ groups, and each group is supposed to utilize one subcarrier; for example, the signals of the users in group $j$ will be multiplexed and transmitted on the $j$th subcarrier. We denote the user grouping results as $\mathcal{I} = \{\mathcal{I}_1,\, \mathcal{I}_2, ...,\mathcal{I}_J\}$, where $\mathcal{I}_j$, given $j\in\mathcal{J}=\{1,...,J\}$, denotes the user set in group $j$ that satisfies $\bigcup_{j=1}^J{\mathcal{I}_j=\mathcal{K}=\{1,...,K\}}$. We let $I_j = |\mathcal{I}_j|$ that satisfies $\sum_{j=1}^J I_j = K$ denote the number of users in group $j$. Note that $I_j\in[0, K]$ is not a fixed value and that $I_j=0$ indicates no signals being transmitted on subcarrier~$j$.

The QoS requirements of URLLC are described as the latency threshold $D^{\mathrm{th}}$ and the overall error probability threshold $\varepsilon^{\mathrm{th}}$ \cite{ref5}. If the channel blocklength (channel uses) is used to describe the latency, then, the transmission should be accomplished within $N^{\mathrm{th}}$ symbols, where $N^{\mathrm{th}} = D^{\mathrm{th}}/T_{\mathrm{s}}$ is the total channel blocklength and $T_{\mathrm{s}} = 1/B$ is the symbol duration \cite{ref18}. Hence, the available channel blocklength of each subcarrier is determined as $N_j = N^{\mathrm{th}}/J$.

\subsection{Channel Model}
As per the relevant protocols, the end-to-end (E2E) latency is less than 1 ms in most URLLC applications, which is less than the channel coherence time, which means that the channel is quasi-static, i.e., the channel conditions remain constant during transmission \cite{add1}. In addition, considering the limited channel feedback in URLLC, the perfect channel estimation is difficult to achieve. Thus, imperfect CSIT and perfect CSIR are assumed in this work. Denote $\mathbf{h}_{j, k} = \sqrt{\alpha_{j,k}} \mathbf{g}_{j,k} \in \mathbb{C}^{M_t \times 1}$, given $j\in\mathcal{J}$ and $k\in\mathcal{I}_j$, as the channel vector between the BS and user $k$ in group $j$, where $\alpha_{j,k}$ is the large-scale fading, and $\mathbf{g}_{j,k}$ is the small-scale Rayleigh fading vector whose elements are modeled as independently and identically distributed (i.i.d.) complex Gaussian variables with zero mean and unit variance. By applying the minimum mean square error (MMSE) estimator, the channel vector can be written as
\begin{equation}\label{R1}
\mathbf{h}_{j,k} = \sqrt{\alpha_{j,k}}(\widehat{\mathbf{g}}_{j, k} + \widetilde{\mathbf{g}}_{j, k}),
\end{equation}
where $\widehat{\mathbf{g}}_{j, k} \in \mathcal{CN}(0, (1-\sigma_e^2)\mathbf{I_{M_t}})$ is the estimated small-scale fading, while $\widetilde{\mathbf{g}}_{j, k}\in \mathcal{CN}(0, \sigma_e^2\mathbf{I_{M_t}})$ quantifies the estimation error with uncertainty $\sigma_e^2=[0,1]$, and they are uncorrelated \cite{ref41}. Apart from that, we consider a time-division-duplex (TDD) system, where the channel estimation is completed in the uplink transmission stage, and the estimated channel will be used for the downlink transmission in a coherent block. Therefore, the model given in (\ref{R1}) reflects the imperfection due to the finite pilot sequence length \cite{refadd1}.

\subsection{Signal Transmission and Reception}
According to the principle of downlink RSMA \cite{ref6}, signal $W_{j, k}$ intended for user $k$ in group $j$ is divided into two parts as $W_{j,k} = W_{j, k,c} + W_{j, k,p}$, where $W_{j, k, c}$ is the common part that is jointly encoded into a common stream $s_{j, c}$, and $W_{j,k,p}$ is the private part that is independently encoded into private streams $s_{j, k}$. Therefore, there are $I_j$ streams, i.e., $\mathbf{s}_j = [s_{j, 1}, ..., s_{j, I_j}, s_{j, c}]^T$, needed to be transmitted on the subcarrier $j$.

By performing linear precoding on the stream of group $j$, the transmitted signal on subcarrier $j$ is denoted as
\begin{equation}\label{eq2}
\mathbf{x}_j = \sqrt{p_{j,c}}\mathbf{w}_{j,c}s_{j,c} + \sum_{k=1}^{I_j}\sqrt{p_{j,k}}\mathbf{w}_{j,k}s_{j,k},
\end{equation}
where $p_{j,c}$ and $p_{j,k}$ are transmit power of $s_{j,c}$ and $s_{j,k}$ that satisfies $\sum_{k\in\mathcal{I}_j} p_{j,k} + p_{j, c} \le p_{j,\max}$, assuming that $\mathbb{E}[\mathbf{s}_j\mathbf{s}_j^H] = \mathbf{I}$, with $p_{j,\max}$ being the maximum transmit power of subcarrier $j$ and $\mathbf{w}_{j,c}$ and $\mathbf{w}_{j,k}$ being beamforming vectors of $s_{j,c}$ and $s_{j,k}$ with $||\mathbf{w}_{j,c}||^2 = 1$ and $||\mathbf{w}_{j,k}||^2 = 1$, respectively. Thus, there are $I_j+1$ beamforming vectors that need to be determined by the BS for group $j$.
 
To avoid the complex precoding design, the regularized zero-forcing beamforming (R-ZFBF)\footnote{Note that ZFBF is mainly used in the scenario with $M_t \ge I_j$, i.e., underloaded scenario. However, to fully explore the advantages of RSMA, ZFBF can also be used in the overloaded scenario with $M_t < I_j$ in this work.} is adopted in this work for private beamforming vector $\mathbf{w}_{j,k}$ \cite{ref42}. Specifically, as per the available channel, R-ZFBF designs $\mathbf{W}_{j,p} = [\mathbf{w}_{j,1},\mathbf{w}_{j,2},...,\mathbf{w}_{j,I_j}]\in\mathbb{C}^{M_t\times I_j}$  as
\begin{equation}\label{eq3}
\mathbf{W}_{j,p} = \zeta \left(\mathbf{\widehat{G}}_j\mathbf{\widehat{G}}_j^H + \kappa\mathbf{I_{M_t}} \right)^{-1}\mathbf{\widehat{G}}_j,
\end{equation}
where $\mathbf{\widehat{G}}_j = [\mathbf{\widehat{g}}_{j,1}, \mathbf{\widehat{g}}_{j,2},..., \mathbf{\widehat{g}}_{j, I_j}]\in \mathbb{C}^{M_t \times I_j}$ is the estimated channel matrix of group $j$; $\kappa$ is the regularization parameter, and $\zeta$ refers to the normalization scalar to guarantee the total transmit power constraint. Similar to \cite{ref42}, $\mathbf{w}_{j,c} = \sum_{k\in\mathcal{I}_j} \bar{\omega}_{j,k} \mathbf{w}_{j,k}$, where $\bar{\omega}_{j,k}=1/\sqrt{M_tI_j}$, is used for the precoding design of $\mathbf{w}_{j,c}$.

At the receiver, the received signal of user $k$ in group $j$ can be expressed as

\begin{equation}\label{eq4}
y_{j,k}  = \!\sqrt{\alpha_{j,k}}{\mathbf{g}}_{j,k}^{H}(\sqrt{p_{j,c}}\mathbf{w}_{j,c}s_{j,c}+\sum_{k'\in \mathcal{I}_j}{\sqrt{p_{j,k'}}\mathbf{w}_{j,k'}s_{j,k'}})
\end{equation}
where $n_{j,k}\sim\mathcal{CN}(0, \sigma_{j,k}^2)$ is the additive white Gaussian noise (AWGN); for simplicity, we assume that $\sigma_{j,k}^2 = \sigma^2$.

The common stream is seen as the strong signal that needs to be decoded first by all users. Thus, the SINR of user $k$ in group $j$ decoding $s_{j,c}$ is denoted as

\begin{align} \label{eq5}
\gamma _{j,k,c} &=\frac{p_{j,c}\rho _{j,k,c}}{\underset{k'\in \mathcal{I} _j}{\sum}{p_{j,k'}\rho _{j,k,k'}+\sigma _{e}^{2}\alpha _{j,k}\left( \underset{k'\in \mathcal{I} _j}{\sum}{p_{j,k'}+p_{j,c}} \right) +\sigma ^2}} 
\notag \\
&=\frac{p_{j,c}\rho _{j,k,c}}{\underset{k'\in \mathcal{I} _j}{\sum}{a_{j,k,k'}p_{j,k'}+b_{j,k}p_{j,c}+c}},
\end{align}
where $\rho_{j,k,c} = \alpha_{j,k}|\widetilde{\mathbf{g}}_{j,k}^H\mathbf{w}_{j,c}|^2$, $\rho_{j,k,k'} = \alpha_{j,k}|\widetilde{\mathbf{g}}_{j,k}^H\mathbf{w}_{j,k'}|^2$, $a_{j,k,l} = \rho_{j,k,l} + \sigma_e^2\alpha_{j,k}$, $b_{j,k} = \sigma_e^2\alpha_{j,k}$, and $c = \sigma^2$. Note that in (\ref{eq5}), since only the estimated channel is available at the BS, any received signals related to estimation errors will be treated as noise. According to (\ref{eq1}), the decoding error probability is denoted as
\begin{equation}\label{eq6}
\varepsilon_{k,j,c} = Q\left(f_{j,k,c}(\gamma_{j,k,c}, R_{j,c})\right),
\end{equation}
where $R_{j,c}$ is the transmission rate of the common stream and $f_{j,k,c}$ is given by
\begin{equation}\label{R2}
f_{j,k,c}(\gamma_{j,k,c}, R_{j,c}) = \ln2\sqrt{\frac{N_j}{V_{j,k,c}}}\left(\log_2(1+\gamma_{j,k,c})-R_{j,c}\right),
\end{equation}
To guarantee the decoding error probability, we have $R_{j,c} \le \underset{k\in\mathcal{I}_j}{\min}\{R_{j,k,c}^{\mathrm{th}}\}$, where $R_{j,k,c}^{\mathrm{th}}$ is given by
\begin{equation}\label{eq7}
R_{j,k,c}^{\mathrm{th}} = \log_2(1+ \gamma_{j,k,c}) - \frac{Q^{-1}(\varepsilon^{\mathrm{th}})}{\ln2}\sqrt{\frac{V_{j,k,c}}{N_j}}.
\end{equation}

If $s_{j,c}$ is successively decoded by user $k$, then, the SIC is performed to subtract $s_{j,c}$. However, due to the imperfect CSIT, the SIC operations cannot be perfectly utilized to subtract $s_{j,c}$ at the user side. Thus, the corresponding SINR and decoding error probability are respectively expressed as
\begin{equation}\label{eq8}
	\gamma_{j,k,p} = \frac{p_{j,k}\rho_{j,k,k}}{\underset{k'\in\mathcal{I}_j \setminus k}{\sum}a_{j,k,k'}p_{j,k'}+b_{j,k}(p_{j,k}+p_{j,c}) +c},
\end{equation}
and
\begin{equation}\label{eq9}
	\varepsilon_{k,j,p} = Q\left(f_{j,k,p}(\gamma_{j,k,p}, R_{j,k,p}) \right),
\end{equation}
where $R_{j,k,p}$ is the transmission rate of the private stream. Besides, due to decoding error probability $\varepsilon_{j,k,c}$, the successful decoding of $s_{j,c}$ cannot be guaranteed \cite{ref18}.  However, we have $W_{j,k}=W_{j,k,c}+W_{j,k,p}$ in URLLC; thus, if user $k$ fails to decode $s_{j,c}$, we can stipulate that the decoding procedure of user $k$ is failed.
\subsection{Problem Formulation}
In this work, we use ET as the performance indicator. For user $k$, the achievable ET for the common stream is defined as
\begin{equation}\label{eq12}
T_{j,k,c} = R_{j,k,c}\left(1 - \varepsilon_{j,k,c}\right),
\end{equation}
where $R_{j,k,c}$ is the common rate allocated to user $k$, which satisfies $\sum_{k\in\mathcal{G}_j}R_{j,k,c}\le R_{j,c}$. Besides, the achievable ET for the private stream is defined as
\begin{align}\label{eq13}
T_{j,k,p} &= R_{j,k,p}(1-\varepsilon_{j,k,c} - (1-\varepsilon_{j,k,c})\varepsilon_{j,k,p}) \notag \\
& \approx R_{j,k,p}(1-\varepsilon_{j,k,p}-\varepsilon_{j,k,c}).
\end{align}
The above approximation is accurate since we require that both $\varepsilon_{j,k,c}$ and $\varepsilon_{j,k,p}$ are not higher than a stringent threshold $\varepsilon^{\mathrm{th}} \le 10^{-5}$ in URLLC \cite{ref20,ref36}.
By synthesizing (\ref{eq12}) and (\ref{eq13}), the achievable ET for subcarrier $j$ can be determined as $T_j=\sum_{k\in\mathcal{I}_j}(T_{j,k,c} + T_{j,k,p})$, and the ET of the system can be determined as $T^{\mathrm{tot}}=\sum_{j\in \mathcal{J}}T_j$.

We aim to maximize the total ET of all subcarriers; thus, the optimization problem is mathematically formulated as follows:
\begin{IEEEeqnarray}{cl}\label{eq14}
	\hspace{-18pt}\max_{\mathcal{I},\mathbf{R}_{j,c},\mathbf{R}_{j,p},\mathbf{p}_j, \forall j} \quad & T^{\mathrm{tot}}=\sum_{j=1}^{J}\sum_{k=1}^{I_j} (T_{j,k,c} + T_{j,k,p}), \IEEEyesnumber  \\
	\text{s.t.}    
	& \sum_{k\in\mathcal{I}_j}R_{j,k,c}\le R_{j,c}, \forall j, \IEEEyessubnumber \label{14a} \\
	& R_{j,c}\le \underset{k\in\mathcal{I}_j}{\min}\{R_{j,k,c}^{\mathrm{th}}\}, \forall j, \IEEEyessubnumber \label{14b} \\
	& R_{k,\min}\le R_{j,k,c} + R_{j,k,p}, \forall j,\forall k\in \mathcal{I}_j, \IEEEyessubnumber \label{14c} \\
	& \varepsilon_{j,k,c}\le \varepsilon^{\mathrm{th}}, \varepsilon_{j,k,p} \le \varepsilon^{\mathrm{th}}, \forall j, \forall k \in \mathcal{I}_j, \IEEEyessubnumber \label{14d} \\
	& \sum_{k\in\mathcal{I}_j}p_{j,k} + p_{j,c} \le p_{j,\max}, \forall j, \IEEEyessubnumber \label{14e}\\
	& \sum_{j\in\mathcal{J}}p_{j,\max} \le P_{\max}, \IEEEyessubnumber \label{14f} \\
	& \mathbf{R}_{j,c}\ge0, \mathbf{R}_{j,p}\ge0,\mathbf{p}_{j}\ge0, \forall j, \IEEEyessubnumber \label{14g}
\end{IEEEeqnarray}
where $\mathbf{R}_{j,c}=[R_{j,1,c},...,R_{j,I_j,c},R_{j,c}]^T\in \mathbb{R}^{(I_j+1)\times 1}$, $\mathbf{R}_{j,p}=[R_{j,1,p},...,R_{j,I_j,p}]^T\in \mathbb{R}^{I_j\times 1}$, and $\mathbf{p}_j=[p_{j,1},...,p_{j,I_j}, p_{j,c}, p_{j, \max}]^T\in \mathbb{R}^{(I_j+2)\times 1}$. In addition, the common rate constraints are represented by (\ref{14a}) and (\ref{14b}). Constraint~(\ref{14c}) signifies the minimum rate requirement, and constraints~(\ref{14d}) regulate the reliability requirement of URLLC. (\ref{14f}) and (\ref{14g}) impose power constraints.

\subsection{Problem Transformation}
Problem~(\ref{eq14}) is, in essence, a mixed-integer non-convex problem since the user grouping, rate control, and power allocation schemes need to be jointly considered. In this regard, we first approximate $T_{j,k,c}$ and $T_{j,k,p}$ using the following lemma.

\par \textit{Lemma 1:} When $\gamma_{j,k,p}\le 2^{91} - 1$ for a given $\varepsilon^{\mathrm{th}} \le 10^{-4}$ and $N_j \le 10^{4}$, the $T_{j,k,c}$ and $T_{j,k,p}$ can be approximated by the following lower bounds:
\begin{equation}
\left\{\begin{array}{l}
T_{j,k,c} \ge \overline{T}_{j,k,c} = R_{j,k,c}(1 - \varepsilon^{\mathrm{th}}), \\
T_{j,k,p} \ge \overline{T}_{j,k,p} = R_{j,k,p}(1 - 2\varepsilon^{\mathrm{th}}).
\end{array}\right.
\end{equation}
\par \textit{Proof:} See Appendix. $\hfill\square$

In URLLC, the typical QoS requirements are $\varepsilon^{\mathrm{th}} \le 10^{-5}$ and $N^{\mathrm{th}} \le 10^{3}$ ($B = 1$ MHz) \cite{ref5}. It is practical to apply Lemma 1 to resource management in stringent URLLC services since $\gamma_{j,k,p}\le 2^{91}-1$ can be guaranteed.

With this approximation, the utility function of (\ref{eq14}) can be rewritten as a liner function as follows:
\begin{equation}\label{eq15}
\overline{T}^{\mathrm{tot}} = \sum_{j\in \mathcal{J}}\sum_{k\in \mathcal{I}_j}(R_{j,k,c}(1 - \varepsilon^{\mathrm{th}})+R_{j,k,p}(1 - 2\varepsilon^{\mathrm{th}})),
\end{equation}
and the problem (\ref{eq14}) can be reformulated as
\begin{IEEEeqnarray}{cl}\label{p2}
	\hspace{-18pt}\max_{\mathcal{I},\mathbf{R}_{j,c},\mathbf{R}_{j,p},\mathbf{p}_j, \forall j} \quad & \overline{T}^{\mathrm{tot}}, \IEEEyesnumber  \\
	\text{s.t.}    
	& R_{j,k,p} \le R_{j,k,p}^{\mathrm{th}}, \forall j, \forall k\in \mathcal{I}_j, \IEEEyessubnumber \label{p2a}\\
	& \text{(\ref{14a})-(\ref{14c})}, \text{(\ref{14e})-(\ref{14g})}. \nonumber
\end{IEEEeqnarray}
However, problem (\ref{p2}) is still a mixed-integer non-convex problem and is intuitively intractable due to the tightly coupled variables. To this end, we use a decomposition method to solve problem~(\ref{p2}). The detailed process is presented in the following section.

\section{The Decomposition Method}\label{Section III}
In this section, we use a decomposition method to address problem~(\ref{p2}). Specifically, we split (\ref{p2}) into two subproblems, i.e., the joint rate control and power allocation subproblem and the user grouping subproblem. Then, we solve these two subproblems separately to obtain a suboptimal solution.

\subsection{Joint Rate Control and Power Allocation}
For a certain user-grouping scheme, we first focus on finding optimal rate control and power allocation schemes. To begin with, we rewrite the problem as follows:
\begin{IEEEeqnarray}{cl}\label{p3}
	\hspace{-18pt}\max_{\mathbf{R}_{j,c},\mathbf{R}_{j,p},\mathbf{p}_j, \forall j} \quad & \overline{T}^{\mathrm{tot}}, \IEEEyesnumber  \\
	\text{s.t.}    
	& \text{(\ref{14a})-(\ref{14c})}, \text{(\ref{14e})-(\ref{14g})}, \text{(\ref{p2a})}. \nonumber
\end{IEEEeqnarray}
It it evident that problem formulated above in (\ref{p3}) is non-convex due to the non-convexity of constraints (\ref{14b}) and (\ref{p2a}). To this end, we propose an iteration-based CCCP method detailed to address (\ref{p3}).
\subsubsection{CCCP Method}
Regarding $\gamma_{j,k,p}$ as a variable, the second-order derivative of $\phi_{j,k,p}(\gamma_{j,k,p})=\sqrt{V_{j,k,p}} = \sqrt{1 - (1+\gamma_{j,k,p})^{-2}}$ w.r.t. $\gamma_{j,k,p}$ is obtained as
\begin{equation}\label{eq17}
\phi_{j,k,p}''(\gamma_{j,k,p}) = -\frac{3V_{j,k,p}(1+\gamma_{j,k,p})^2+1}{V_{j,k,p}^{3/2}(1+\gamma_{j,k,p})^6} \le 0.
\end{equation}
Thus, $V_{j,k,p}$ is a concave function of $\gamma_{j,k,p}$ (the same conclusion applies to $\phi_{j,k,c}(\gamma_{j,k,c}) = \sqrt{V_{j,k,c}}$), which means that both rate functions $R_{j,k,c}^{\mathrm{th}}$ and $R_{j,k,p}^{\mathrm{th}}$ are the difference between two concave functions w.r.t. $\gamma_{j,k,p}$. In this regard, the CCCP method is eligible to be deployed to solve this problem.

In particular, we first introduce slack variables $\bm{\lambda}_j = [\lambda_{j,1},...,\lambda_{j,I_j}]^T\in \mathbb{R}^{I_j\times 1}$ and $\bm{\mu}_j = [\mu_{j,1},...,\mu_{j,I_j}]^T\in \mathbb{R}^{I_j\times 1}$ to rewrite rate functions $R_{j,k,c}^{\mathrm{th}}$ and $R_{j,k,p}^{\mathrm{th}}$ as follows:
\begin{IEEEeqnarray}{cl}\label{p4}
\underset{\begin{array}{c} \mathbf{R}_{j,c},\mathbf{R}_{j,p},\mathbf{p}_j, \\ \bm{\lambda}_j, \bm{\mu}_j \end{array}}{\max} &  \sum_{j\in\mathcal{J}}\sum_{k\in \mathcal{I}_j} \overline{T}^{\mathrm{tot}}, \IEEEyesnumber  \\
	\text{s.t.}    
	& R_{j,c} \le \underset{k\in\mathcal{I}_j}{\min}\{R_{j,k,c}^{\mathrm{th}}(\lambda_{j,k})\}, \forall j, \IEEEyessubnumber \label{p4a} \\
	& R_{j,k,p} \le R_{j,k,p}^{\mathrm{th}}(\mu_{j,k}), \forall j, \forall k\in \mathcal{I}_j, \IEEEyessubnumber \label{p4b} \\
	& \lambda_{j,k} \le \gamma_{j,k,c}, \, \forall j, \forall k \in \mathcal{I}_j, \IEEEyessubnumber \label{p4c1} \\
	& \mu_{j,k} \le \gamma_{j,k,p}, \forall j, \forall k \in \mathcal{I}_j, \IEEEyessubnumber \label{p4c2} \\
	& \bm{\lambda}_j \ge 0, \bm{\mu}_j \ge 0, \forall j, \IEEEyessubnumber \label{p4d}\\
	& (\text{\ref{14a}}), (\text{\ref{14c}}),\text{(\ref{14e})-(\ref{14g})}. \nonumber
\end{IEEEeqnarray}
Note that the new non-convex constraints (\ref{p4c1}) and (\ref{p4c2}) are further introduced in the above problem. For non-convex constraints (\ref{p4a}) and (\ref{p4b}), we can linearize $\phi_{1,p}(\mu_{j,k})$ at point $\mu_{j,k}^{[n]}$, where $[n]$ denotes $n$th iteration of the CCCP method, as follows:
\begin{align}\label{eq19}
\phi_{j,k,p}(\mu_{j,k}) & \le \phi_{j,k,p}(\mu_{j,k}^{[n]}) + \frac{\mu_{j,k}-\mu_{j,k}^{[n]}}{\phi_{j,k,p}(\mu_{j,k}^{[n]})(1 + \mu_{j,k}^{[n]})^3} \notag \\ &\overset{\triangle}{=}\widetilde{\phi}_{j,k,p}(\mu_{j,k};\mu_{j,k}^{[n]}).
\end{align}
The same conclusion applies to $\phi_{j,k,c}(\lambda_{j,k})$ as follows:
\begin{align}\label{add1}
\phi_{j,k,c}(\lambda_{j,k}) & \le \phi_{j,k,c}(\lambda_{j,k}^{[n]}) + \frac{\lambda_{j,k}-\lambda_{j,k}^{[n]}}{\phi_{j,k,c}(\lambda_{j,k}^{[n]})(1 + \lambda_{j,k}^{[n]})^3} \notag \\ &\overset{\triangle}{=}\widetilde{\phi}_{j,k,c}(\lambda_{j,k};\lambda_{j,k}^{[n]}).
\end{align}
Hence, by substituting $\phi_{j,k,p}(\mu_{j,k})$ with $\widetilde{\phi}_{j,k,p}(\mu_{j,k};\mu_{j,k}^{[n]})$ and $\phi_{j,k,c}(\lambda_{j,k})$ with $\widetilde{\phi}_{j,k,c}(\lambda_{j,k};\lambda_{j,k}^{[n]})$, constraints (\ref{14b}) and (\ref{p2a}) can be approximated as convex alternatives, respectively, as follows:
\begin{equation}\label{eq20}
R_{j,c} \le \min_{k\in\mathcal{I}_j}\left\{\log_2(1 + \lambda_{j,k}) - \theta\widetilde{\phi}_{j,k,c}(\lambda_{j,k};\lambda_{j,k}^{[n]})\right\},
\end{equation}
\begin{equation}\label{eq21}
R_{j,k,p}^{\mathrm{th}} \le \log_2(1 + \mu_{j,k}) -\theta\widetilde{\phi}_{j,k,p}(\mu_{j,k};\mu_{j,k}^{[n]}),
\end{equation}
where $\theta=\frac{Q^{-1}(\varepsilon^{\mathrm{th}})}{\sqrt{N_j}\ln(2)}$.

Next, we focus on the newly introduced inequality $\lambda_{j,k,c} \le \gamma_{j,k,c}$, which can be expressed as

\begin{align}\label{eq22}
& \varLambda_{j,k,c}\left( \bm{\lambda}_{j}, \mathbf{p}_j \right)
\notag \\
 =&\sum_{k'\in \mathcal{I} _j}{a_{j,k,k'}p_{j,k'}\lambda _{j,k}+b_{j,k}p_{j,c}\lambda _{j,k}+c\lambda _{j,k}}
\notag \\
=&\sum_{k'\in \mathcal{I} _j}{\frac{a_{j,k,k'}}{4}\left[ \left( p_{j,k'}+\lambda _{j,k} \right) ^2-\left( p_{j,k'}-\lambda _{j,k} \right) ^2 \right]}
\notag \\
&+\frac{b_{j,k}}{4}\left[ \left( p_{j,c}+\lambda _{j,k} \right) ^2-\left( p_{j,c}-\lambda _{j,k} \right) ^2 \right] +c\lambda _{j,k}
\notag \\
\le& \rho_{j,k,c}p_{j,c}.
\end{align}
Obviously, the left-hand side of (\ref{eq22}) is a DC function. We thus let $\xi(x, y)=(x - y)^2$, at point $(x^{[n]},y^{[n]})$, $\xi(x, y)$ can be linearized as
\begin{align}\label{eq23}
\xi(x, y) &\ge -( x^{[n]}-y^{[n]}) ^2+2( x^{[n]}-y^{[n]}) \left( x-y \right) 
\notag \\
&\triangleq \widetilde{\xi }(x, y).
\end{align}
With (\ref{eq23}), $\varLambda _{j,k,c}\left( \bm{\lambda _{j}},\mathbf{p}_j \right)$ can be approximated as
\begin{align}\label{eq24}
& \varLambda _{j,k,c}\left( \bm{\lambda _{j}},\mathbf{p}_j \right)
\notag \\
 \le& \sum_{k'\in \mathcal{I} _j}{\frac{a_{j,k,k'}}{4}\left[ \left( p_{j,k'}+\lambda _{j,k} \right) ^2-\widetilde{\xi }(p_{j,k'}, \lambda_{j,k}) \right]}
\notag \\
&+\frac{b_{j,k}}{4}\left[ \left( p_{j,c}+\lambda _{j,k} \right) ^2- \widetilde{\xi}(p_{j,c}, \lambda_{j,k}) \right] +c\lambda _{j,k}
\notag \\
\triangleq& \widetilde{\varLambda}_{j,k,c}\left( \bm{\lambda}_{j},\mathbf{p}_j; \bm{\lambda}_{j}^{[n]}, \mathbf{p}_j^{[n]}\right) 
\end{align}
Therefore, the inequality $\lambda_{j,k,c} \le \gamma_{j,k,c}$ can be approximated by the following convex form:
\begin{equation}\label{eq25}
\widetilde{\varLambda}_{j,k,c}\left( \bm{\lambda}_{j},\mathbf{p}_j; \bm{\lambda}_{j}^{[n]}, \mathbf{p}_j^{[n]}\right)  \le \rho_{j,k,c}p_{j,c},
\end{equation}
By utilizing the similar steps in (\ref{eq22})-(\ref{eq25}), the inequality $\mu_{j,k,p}\le \gamma_{j,k,p}$ can be approximated as
\begin{equation}\label{eq26}
\widetilde{\varLambda}_{j,k,p}\left( \bm{\mu}_{j},\mathbf{p}_j; \bm{\mu}_{j}^{[n]}, \mathbf{p}_j^{[n]}\right)  \le \rho_{j,k,k}p_{j,k},
\end{equation}
where
\begin{align}\label{eq27}
	&\widetilde{\varLambda}_{j,k,p}\left( \bm{\mu}_{j},\mathbf{p}_j; \bm{\mu}_{j}^{[n]}, \mathbf{p}_j^{[n]}\right) 
	\notag \\
	=& \sum_{k'\in \mathcal{I} _j\setminus k}{\frac{a_{j,k,k'}}{4}\left[ \left( p_{j,k'}+\mu _{j,k} \right) ^2-\widetilde{\xi }(p_{j,k'}, \mu_{j,k}) \right]}
	\notag \\
	&+\frac{b_{j,k}}{4}\left[ \left( p_{j,c}+\mu _{j,k} \right) ^2- \widetilde{\xi}(p_{j,c}, \mu_{j,k}) \right]
	\notag \\
	& + \frac{b_{j,k}}{4}\left[ \left( p_{j,k}+\mu _{j,k} \right) ^2- \widetilde{\xi}(p_{j,k}, \mu_{j,k}) \right] +c\mu _{j,k}
\end{align}
Therefore, at the $n$th CCCP iteration, problem~(\ref{p4}) can be transformed into a convex problem, as follows:
\begin{IEEEeqnarray}{cl}\label{eq28}
\underset{\begin{array}{c} \mathbf{R}_{j,c},\mathbf{R}_{j,p},\mathbf{p}_j, \\ \bm{\lambda}_j, \bm{\mu}_j \end{array}}{\max} &  \sum_{j\in\mathcal{J}}\sum_{k\in \mathcal{I}_j} \overline{T}^{\mathrm{tot}}, \IEEEyesnumber  \\
	\text{s.t.}    
	&(\text{\ref{eq20}}),\,(\text{\ref{eq21}}) ,\, (\text{\ref{eq25}}), \, (\text{\ref{eq26}}),\,(\text{\ref{14a}}),\, \notag \\ &(\text{\ref{14c}}),\,(\text{\ref{14e}})-(\text{\ref{14g}}),\, \text{(\ref{p4d})}. \IEEEnonumber
\end{IEEEeqnarray}
Owing to its convexity, problem~(\ref{eq28}) can be effectively addressed using the MATLAB CVX toolbox. As shown in Algorithm \ref{algorithm 1}, the suboptimal power allocation results are obtained by repeatedly solving problem~(\ref{eq28}) until convergence is achieved. In addition, the convergence to the local optimality has been proven to be achievable by CCCP as given in \cite{ref43}.

\begin{algorithm}[t]
	\caption{CCCP for the problem (\ref{p3})}\label{algorithm 1}
	\LinesNumbered
	Initialize $\bm{\lambda}_j^{[0]}$, $\bm{\mu}_j^{[0]}$, $\mathbf{p}_j^{[0]}, \forall j$, and tolerance $\varepsilon$;\\
    Set $\overline{T}^{\mathrm{tot}}[n] = 0$ and iteration index $n=0$; \\
	\Repeat{$|\overline{T}^{\mathrm{tot}}[n]-\overline{T}^{\mathrm{tot}}[n-1]| \le \varepsilon$}
	{
		Let $n \gets n+1$;\\
		Solve convex problem~(\ref{eq28}) according to $[\bm{\lambda}_j^{[n-1]}, \bm{\mu}_j^{[n-1]}, \mathbf{p}_j^{[n-1]}]$;\\
		Record optimal solution and value to (\ref{eq28}) by $\mathbf{S}_j^{[n]} \gets [\bm{\lambda}_j^{[n]}, \bm{\mu}_j^{[n]}, \mathbf{p}_j^{[n]}]$ and $\overline{T}^{\mathrm{tot}}[n]$;\\
	}
	Return the local optimal solution to (\ref{p3}).
\end{algorithm}

\subsubsection{LBA Method}
The above proposed CCCP method can effectively solve problem~(\ref{p3}). But the one-layer iteration of CCCP leads to high complexity and thereby running time. To this end, in this subsection, we propose an iteration-free LBA method. Specifically, since $\phi_{j,k,c} = \sqrt{V_{j,k,c}} \le 1$, by letting $\phi_{j,k,c}=1$, the lower bound on $R_{j,k,c}^{\mathrm{th}}$ is obtained as\footnote{Note that $V= 1 - (1+\gamma)^{-2}\approx 1$ when $\gamma\ge 5$ dB; thus, the approximation $V=1$ is accurate in the high SINR region \cite{ref40}.}
\begin{equation}\label{eq29}
\overline{R}_{j,k,c}^{\mathrm{th}} = \log_2(1+\gamma_{j,k,c}) - \theta.
\end{equation}
The same applies to $\overline{R}_{j,k,p}^{\mathrm{th}}$. Therefore, the constraints (\ref{14b}) and (\ref{p2a}) are guaranteed if we replace $R_{j,k,c}^{\mathrm{th}}$ and $R_{j,k,p}^{\mathrm{th}}$ with $\overline{R}_{j,k,c}$ and $\overline{R}_{j,k,p}$, respectively.

However, even with the above approximation, solving problem~(\ref{p3}) is still a non-trivial task due to linear fractional expressions of $\gamma_{j,k,c}$ and $\gamma_{j,k,p}$. Considering this, we first assume that $p_{j,\max} = P_{\max}/J$ and then use equivalent infinitesimal theory to further approximate $\overline{R}_{j,k,c}^{\mathrm{th}}$ and $\overline{R}_{j,k,p}^{\mathrm{th}}$. In specific, according to ($\ref{eq8}$), $R_{j,k,p}^{\dag}$ can be approximated as
\begin{align}\label{eq30}
&\overline{R}_{j,k,p}^{\mathrm{th}} 
\notag \\
=&  \log_2(1 + \gamma_{j,k,p}) -\theta 
\notag \\
\ge & \log _2\left(\!1 \!+\!\frac{p_{j,k}\rho _{j,k,k}}{\underset{k'\in\mathcal{I}_j\setminus k}{\sum}\!\!{p_{j,k'}\rho _{j,k,k'}+b_{j,k}p_{j,\max}+c}} \right)  \!\!- \!\theta 
\notag \\
= & \log _2\left( \sum_{k'\in\mathcal{I}_j}{p_{j,k'}\rho _{j,k,k'}+b_{j,k}p_{j,\max}+c}\right)  \notag \\
   & - \log _2\left( \sum_{k'\in\mathcal{I}_j \setminus k}{p_{j,k'}\rho _{j,k,k'}+b_{j,k}p_{j,\max}+c}\right) - \theta
\notag \\
\overset{\triangle}{=} & \chi_{j,k,p}^{(1)}(\mathbf{p}_j) - \chi_{j,k,p}^{(2)}(\mathbf{p}_j) - \theta;
\end{align}
Furthermore, $\phi_{3,p}^{(2)}(\mathbf{p}_j)$ can be approximated as
\begin{align}\label{eq31}
\chi_{j,k,p}^{(2)}(\mathbf{p}_j) &\!=\! \ln\left( 1+\frac{\underset{k'\in \mathcal{I}_j \setminus k}{\sum}{p_{j,k'}\rho _{j,k,k'}}}{z_{j,k}} \right)/\ln(2) \!+\! \log _2\left( z_{j,k} \right) \notag \\
& \overset{(a)}{\le} \frac{\underset{k'\in \mathcal{I} _j \setminus k}{\sum}{p_{j,k'}\rho _{j,k,k'}}}{z_{j,k}\ln(2)} +\log _2\left( z_{j,k}\right) \notag \\
& \overset{\triangle}{=} \widetilde{\chi}_{j,k,p}^{(2)}(\mathbf{p}_j) ,
\end{align}
where $z_{j,k} = b_{j,k}p_{j,\max} + c$ and (a) is precise because $\ln(1+x)\approx x$ when $-1\leq x\leq 1$ and $\sum_{k'\in \mathcal{I} _j \setminus k}{p_{j,k'}\rho _{j,k,k'}} \approx 0$ when $M_t > K$ with ZFBF. By substituting $\widetilde{\chi}_{j,k,p}^{(2)}(\mathbf{p}_j)$ back into (\ref{eq30}), we have
\begin{equation}\label{eq32}
\overline{R}_{j,k,p}^{\mathrm{th}} \ge \chi_{j,k,p}^{(1)}(\mathbf{p}_j) - \widetilde{\chi}_{j,k,p}^{(2)}(\mathbf{p}_j) - \theta \overset{\triangle}{=} \widehat{R}_{j,k,p}^{\mathrm{th}}.
\end{equation}
	
Similarly, according to (\ref{eq5}), the $R_{j,k,c}^{\dag}$ is approximated as
\begin{align}\label{eq33}
& \overline{R}_{j,k,c}^{\mathrm{th}}  \notag \\
\ge  & \log _2\left(\!1 \!+\!\frac{p_{j,c}\rho _{j,k,c}}{\underset{k'\in\mathcal{I}_j}{\sum}\!\!{p_{j,k'}\rho _{j,k,k'}+b_{j,k}p_{j,\max}+c}} \right)  \!\!- \!\theta \notag \\
\ge & \log _2\left( \sum_{k'\in\mathcal{I}_j}{p_{j,k'}\rho _{j,k,k'}+b_{j,k}p_{j,\max}+ p_{j,c}\rho_{k,c} + c}\right)  \notag \\
	& - \log _2\left( \sum_{k'\in\mathcal{I}_j}{p_{j,k'}\rho _{j,k,k'}+b_{j,k}p_{j,\max}+c}\right) - \theta \notag \\
	\overset{\triangle}{=} & \chi_{j,k,c}^{(1)}(\mathbf{p}_j) - \chi_{j,k,p}^{(1)}(\mathbf{p}_j) - \theta.
\end{align}
By synthesizing (\ref{eq33}) and (\ref{eq32}), $\overline{R}_{j,k,c}^{\mathrm{th}}$ can be expressed as
\begin{equation}\label{eq34}
\overline{R}_{j,k,c}^{\mathrm{th}} \ge \chi_{j,k,c}^{(1)}(\mathbf{p}_j) - \widehat{R}_{j,k,p}^{\mathrm{th}} - \widetilde{\chi}_{j,k,p}^{(2)}(\mathbf{p}_j) - 2\theta \overset{\triangle}{=} \widehat{R}_{j,k,c}^{\mathrm{th}}.
\end{equation}
It is evident that $\widehat{R}_{j,k,c}^{\mathrm{th}}$ is a concave function if we employ $\widehat{R}_{j,k,p}^{\mathrm{th}}$ as a slack variable.

By applying the above approximations, problem~(\ref{p2}) can be reformulated as follows:
\begin{IEEEeqnarray}{cl}\label{eq35}
	\!\!\!\! \underset{\mathbf{R}_{j,c},\mathbf{R}_{j,p},\widehat{\mathbf{R}}_{j,p},\mathbf{p}_j}{\max}&  \quad   \overline{T}^{\mathrm{tot}}, \IEEEyesnumber  \\
	\text{s.t.}&
	 \widehat{R}_{j,k,p}^{\mathrm{th}} \!\le\! \chi_{j,k,p}^{(1)} \!-\! \widetilde{\chi}_{j,k,p}^{(2)}\!-\! \theta, \forall j, \forall k \!\in\! \mathcal{I}_j, \IEEEyessubnumber \label{35a} \\
	& R_{j,c}\le \underset{k\in\mathcal{I}_j}{\min}\{\widehat{R}_{j,k,c}^{\mathrm{th}}\}, \forall j, \IEEEyessubnumber \label{35b} \\
	& R_{j,k,p} \le \widehat{R}_{j,k,p}^{\mathrm{th}}, \forall j, k\in \mathcal{I}_j, \IEEEyessubnumber \label{35c} \\
	& \widehat{\mathbf{R}}_{j,p} \ge 0, \forall j,  \IEEEyessubnumber \label{35d} \\
	&(\text{\ref{14a}}),\,(\text{\ref{14c}}), \,(\text{\ref{14e}}),\,(\text{\ref{14g}}),\IEEEnonumber
\end{IEEEeqnarray}
where $\widehat{\mathbf{R}}_{j,p}= [\widehat{R}_{j,1,p}^{\mathrm{th}},...,\widehat{R}_{j,I_j,p}^{\mathrm{th}}]^T\in \mathbb{R}^{I_j\times 1}$. The convexity of problem~(\ref{eq35}) is guaranteed since $\chi_{j,k,c}^{(1)}(\mathbf{p}_j)$ and $\chi_{j,k,p}^{(1)}(\mathbf{p}_j)$ are concave functions, and $\widetilde{\chi}_{j,k,p}^{(2)}(\mathbf{p}_j)$ is a linear function. Therefore, the CVX toolbox can also be used to solve problem~(\ref{eq35}).

\subsubsection{Feasibility and Complexity Analysis}
Due to the impacts of the imperfect CSIT and the minimum transmission rate requirement, the feasibility of problem~(\ref{p2}) might not necessarily be guaranteed, which is thus worth investigating. However, it is difficult to mathematically prove the feasibility of problem~(\ref{p2}) directly due to its non-convexity. Hence, we only check the feasibility of problem (\ref{p3}), i.e., under the given user-grouping schemes. That means if (\ref{p3}) is infeasible for a given $\mathcal{I}$, $T^{\mathrm{tot}}$ will be set to be 0. Considering that convex problem (\ref{eq35}) is the inner approximation of (\ref{p3}); that is, problem (\ref{p3}) is feasible if (\ref{eq35}) is feasible. In this part, we use the feasibility of (\ref{eq35}) to evaluate the feasibility of (\ref{p3}).

Specifically, the feasibility-checking problem for (\ref{eq35})  can be expressed as
\begin{IEEEeqnarray}{cl}\label{eq36}
	\!\!\!\! \underset{\mathbf{R}_{j,c},\mathbf{R}_{j,p},\widehat{\mathbf{R}}_{j,p},\mathbf{p}_j, r}{\max}&  \quad r, \IEEEyesnumber  \\
	\text{s.t.}
	& r \le R_{j,k,c} + R_{j,k,p}, \forall j, \forall k\in \mathcal{I}_j. \IEEEyessubnumber \\
	&(\text{\ref{14a}}), \text{(\ref{35a})}-\text{(\ref{35d})},\,(\text{\ref{14e}}),\,(\text{\ref{14g}}).\IEEEnonumber
\end{IEEEeqnarray}
Similar to problem~(\ref{eq35}), problem~(\ref{eq36}) is a convex problem that can be effectively solved by the CVX toolbox. By denoting the optimal value of problem~(\ref{eq36}) as $r^{*}$, we can make an observation that problems (\ref{eq35}) as well as (\ref{p3}) are feasible if and only if $r^{*}\ge R_{k,\min}$. Additionally, the power allocation coefficients obtained from (\ref{eq36}) can be utilized to initialize the CCCP procedure if problem~(\ref{p3}) is feasible.

Next, we focus on the computational complexity of the proposed CCCP and LBA methods. To begin with, if the interior point algorithm is used to solve convex problems (\ref{eq28}) and (\ref{eq35}), then the complexity of these two problems is $\mathcal{O}\left((5K+2J)^{3.5}/\log_2(\varepsilon)\right)$ and $\mathcal{O}\left((3K +2J)^{3.5}/\log_2(\varepsilon)\right)$, respectively, where $\varepsilon$ is the approximation error. In addition, since problem~(\ref{eq28}) needs to be iteratively solved via the CCCP method, by denoting $T_1$ as the number of iterations, the complexity of the CCCP method can be determined as $\mathcal{O}\left(T_1(5K+2J)^{3.5}/\log_2(\varepsilon)\right)$. In a similar way, the complexity of the LBA method is $\mathcal{O}\left((3K +2J)^{3.5}/\log_2(\varepsilon)\right)$ since no iteration is required.

\subsection{User Grouping Schemes}
This subsection focuses on finding appropriate user-grouping schemes, i.e., $\mathcal{I}$, that maximizes $T^{\mathrm{tot}}$ in problem (\ref{p2}). In this work, we consider that the arbitrary number of users can be multiplexed on a single subcarrier. Therefore, for the proposed model, the optimal user grouping scheme can only be obtained by an exhaustive search method, which takes up to $J^{K}$ combinations. The complexity of the exhaustive search method grows exponentially with $K$, which is unacceptable in URLLC services. To this end, we first propose a suboptimal greedy search method and then develop a heuristic grouping method.

\subsubsection{Greedy Grouping Method}
Since each user takes up one subcarrier and the number of users multiplexed on a single subcarrier is not restricted, we can design a greedy search method based on the following steps:
\begin{itemize}
\item[\textit{a.}] We initialize set $\mathcal{I}_j = \emptyset$, $\mathcal{I}=\{\mathcal{I}_1,..., \mathcal{I}_J\}$, and $I_j = 0$, $\forall j$, Then we define $S_{j,k}$, where $j\in\mathcal{J}$ and $k\in\mathcal{K}$, as the state of the $k$th user on the $j$th subcarrier. 
\item[\textit{b.}] We let $T^{\mathrm{tot}}\{\mathcal{I}|\mathcal{I}_j \gets S_{j,k}\}$ denote the achievable ET for state $S_{j,k}$, where $\mathcal{I}_j \gets S_{j,k}$ denotes that the user index $k$ has been added to set $\mathcal{I}_j$. It is evident that $T^{\mathrm{tot}}\{\mathcal{I}|\mathcal{I}_j \gets S_{j,k}\}$ depends on both $S_{j,k}$ and $\mathcal{I}$.
\item[\textit{c.}] We initialize the state $S_{j,1}$ according to
\begin{equation}\label{eq38}
\left\{\begin{array}{l}
		j^{\dag} = \underset{j\in \mathcal{J}}{\arg \max}\left\{T^{\mathrm{tot}}\{\mathcal{I}|\mathcal{I}_j\gets S_{j,1}\}\right\} \\
		\mathcal{I}_{j^{\dag}} \gets \{k\}
	\end{array}
 \right.
\end{equation}
Note that $T^{\mathrm{tot}}\{\mathcal{I}|\mathcal{I}_j \gets S_{j,k}\}$ can be obtained through CCCP or LBA method proposed in Subsection A.
\item[\textit{d.}] For each $k$, we update $\mathcal{I}$ according to (\ref{eq38}). The whole process will be completed when $k= K$. After that, all user indexes will be recorded in set $\mathcal{I}$, which will be utilized as the final result for the greedy grouping method.
\end{itemize}

Note that the greedy grouping method can only be used to find a suboptimal solution to the user-grouping problem since it highly depends on the initial order of users. However, this method only contains $JK$ combinations, which is massively less than an exhaustive search.

\subsubsection{Heuristic Grouping Method}
While the greedy grouping method can be used to tackle the user grouping problem with low complexity. The dependence on the rate control and power allocation schemes introduces an additional processing delay in URLLC. In this regard, we further develop a heuristic grouping method to yield insights into the user grouping problem in MC-RSMA systems. The basic idea of the heuristic grouping method is that we tend to schedule users with small user channel correlation coefficients on a subcarrier to achieve better performance of ZFBF. The detailed steps are given as follows:
\begin{itemize}
\item[\textit{a.}]Initialize $\widehat{\mathcal{K}}=\mathcal{K}$ and $\widetilde{\mathcal{K}}=\varnothing$ as the unscheduled user set and scheduled user set, respectively.
\item[\textit{b.}] Define the correlation coefficient between users $k$ and $k'$ as
\begin{equation}\label{eq39}
\Upsilon _{k,k'}=\frac{\parallel \widehat{\mathbf{g}}_{k}^{H}\widehat{\mathbf{g}}_{k'}\parallel}{||\widehat{\mathbf{g}}_k\parallel \parallel \widehat{\mathbf{g}}_{k'}\parallel}.
\end{equation}
It is obvious $\Upsilon_{k,k}=1$ and $\Upsilon_{k,k'}\le 1$, $\forall k' \ne k$. Obtain all possible $\Upsilon_{k,k'}$ via (\ref{eq39}). Then, set $\Upsilon_{k,k}=0$.
\item[\textit{c.}]For all $j \in \{1, 2, ..., J-1\}$, we schedule users on the $j$th subcarrier based on
\begin{equation}\label{eq40}
\left\{\begin{array}{l}
	\mathcal{I}_j \gets \mathcal{K}_j \in \{k'|\Upsilon_{k, k'} \le \Upsilon_{\mathrm{th}}\}, \forall k, k'\in \widehat{\mathcal{K}} \\
	\widehat{\mathcal{K}} = \widehat{\mathcal{K}} \setminus \mathcal{K}_j, \widetilde{\mathcal{K}}=\widetilde{\mathcal{K}} \cup \mathcal{K}_j
	\end{array}\right.,
\end{equation}
where $\Upsilon_{\mathrm{th}}$ is a predefined threshold.
\item[\textit{d.}] Schedule all remaining users in $\widehat{\mathcal{K}}$ on the $J$th subcarrier, i.e., let $\mathcal{I}_J=\widehat{\mathcal{K}}$.
\end{itemize}

It is evident that the performance of the proposed heuristic grouping method depends on threshold $\Upsilon_{\mathrm{th}}$. Therefore, how to determine $\Upsilon_{\mathrm{th}}$ is a crucial problem for this method. To reduce the grouping complexity, we consider the least sophisticated method, that is, finding $\Upsilon_{\mathrm{th}}$ that results in an equal distribution of people across all subcarriers. However, since the number of users $K$ is not necessarily divisible by the number of subcarriers $J$, an appropriate $\Upsilon_{\mathrm{th}}$ needs to guarantee that $I_j=\lfloor K/J\rfloor + i$, where $i=0$ for $r > j$ and $i=1$ otherwise, with $r$ being the remainder of $K$ divided by $J$.

Obviously, the results of the proposed heuristic grouping method are independent of rate control and power allocation schemes. Therefore, using this method will not introduce an additional processing delay for URLLC systems, which indicates that the complexity of the heuristic grouping method is much lower than that of the greedy grouping method.

\section{Simulation Results}\label{Section V}
This section presents simulation results to evaluate the effectiveness and efficiency of the proposed optimization methods. In all simulations, we assume that $K$ users are uniformly distributed in the range between 10 m and 250 m away from the BS, i.e., $d_k \in (10, 300)$. The large-scale path loss model is set by $35.3+37.6\log_{10}(d_k)$ dB and is based only on distance $d_k$. We further configure the total system bandwidth as $B=1$ MHz and the power spectral density of AWGN as -173 dBm/Hz, as well as the noise power as $\sigma = -113$ dBm. In addition, the QoS requirements of URLLC are set as $D^{\mathrm{th}} = 1$ ms, equivalent to $N^{\mathrm{th}} = 1000$, and $\varepsilon^{\mathrm{th}} = 10^{-5}$. The variance of the estimation error due to the imperfect CSIT is set as $\sigma_e^2 = 0.05$. Besides, the minimum rate requirement for each user is set as $R_{k, \min}=1$ bit/s/Hz. The regularization parameter of R-ZFBF is set as $\kappa = I_j\sigma^2$. The above parameters are listed in Table \ref{Table I}.

\begin{figure}[!t]
	\centering
	\includegraphics[width=2.7in]{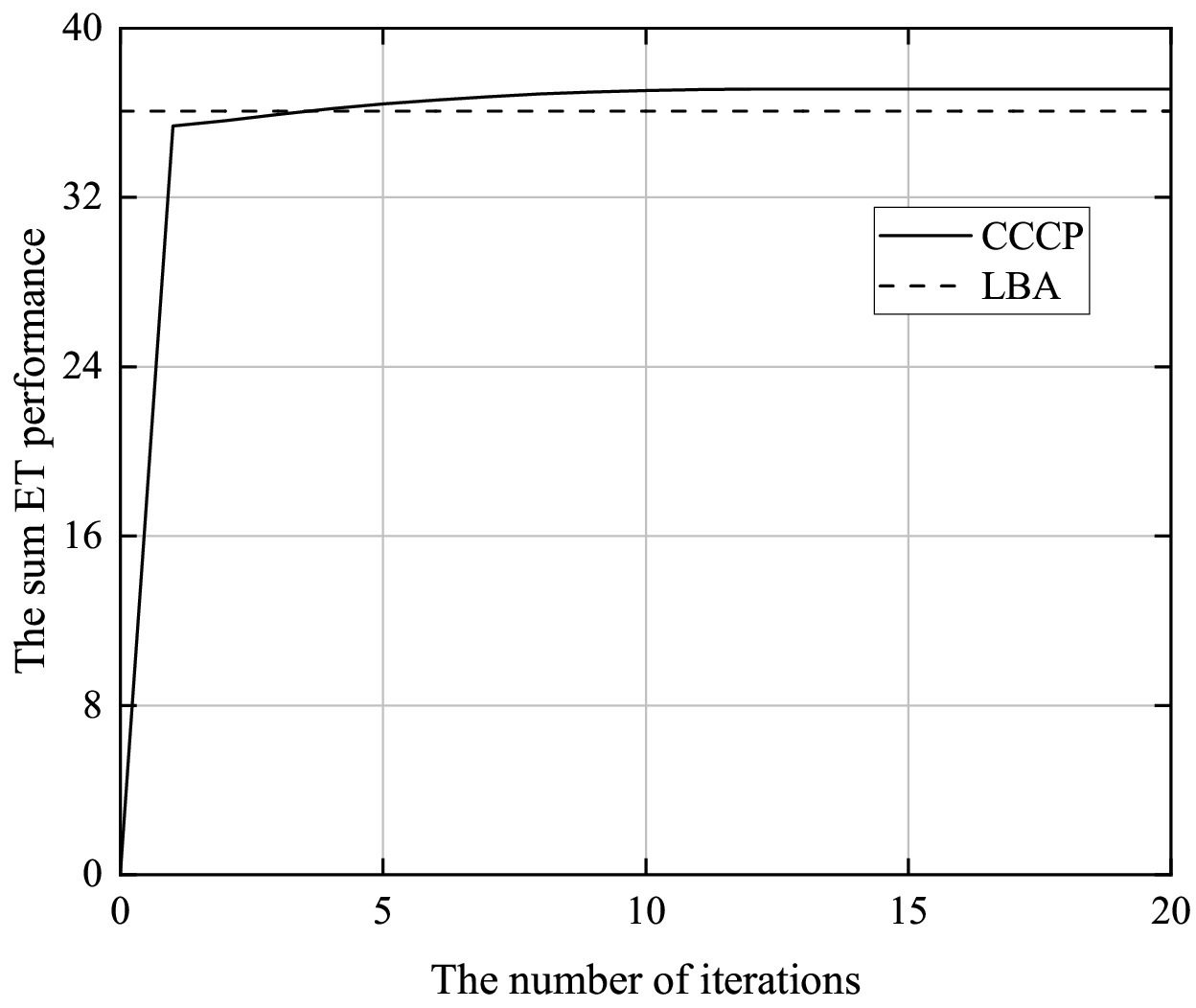}
	\caption{Convergence performance of the CCCP method, given $K=6$ and $J=1$.}
	\label{Convergence}
\vspace{-15pt}
\end{figure}

Furthermore, we evaluate the SDMA scheme to compare its performance with RSMA in URLLC networks. We set power $p_{j,c}=0$ in the simulation to degrade RSMA to SDMA and use the same methods to obtain the corresponding numerical results. All simulations, except Fig. \ref{Convergence}, are the average over 100 independent channel realizations on the MATLAB platform supported by 13th Gen Intel(R) Core(TM) i7-13700 2.10 GHz processor and 32 GB RAM. 

\begin{table}[!t]
	\renewcommand{\arraystretch}{1.3}
	\caption{Simulation parameters.}
	\label{Table I}
	\centering
	\begin{tabular}{|l|l|}
		\hline Parameter                                                       & Value \\
		\hline Distance between users and the BS & $10 - 300$ m \\
		\hline Large-scale fading model                                & $35.3+37.6\lg(d_k)$ dB \\
		\hline Number of transmit antennas	$M_t$				& 32 \\
		\hline Number of users $K$									& 8 \\
		\hline Number of subcarriers $J$							& 3\\
		\hline Power of noise $\sigma^2$                             & -113 dBm \\
		\hline Maximum transmit power $P_{\max}$  	        & 30 dBm\\
		\hline System bandwidth $B$                	      			  & 1 MHz \\
		\hline Minimal rate requirement $R_{k, \min}$              & 1 bit/s/Hz \\
		\hline Error probability threshold $\varepsilon^{\mathrm{th}}$ & $10^{-5}$ \\
		\hline Latency threshold $D^{\mathrm{th}}$ 							& 1 ms \\
		\hline Variance of the estimation error $\sigma_e^2$ & 0.05 \\
		\hline
	\end{tabular}
\end{table}

To begin with, the convergence performance of the proposed CCCP method is evaluated in Fig. \ref{Convergence}, given $K=6$ and $J=1$. The numerical results shown in Fig. \ref{Convergence} are derived based on a single channel realization. It is evident that the CCCP method gradually converges after 10 iterations, which clearly corroborates the convergence of CCCP. Meanwhile, after convergence of the CCCP, e.g., at the 15th iteration, we see that the ET achieved by the CCCP method is 2.78\% higher than that achieved by the LBA method. This observation indicates that the performance gap between the two methods is negligible. Considering that there are no outer-layer iterations required in the proposed LBA method, it is worthwhile to sacrifice a portion of the ET performance to reduce complexity, especially in URLLC scenarios.
\begin{figure}[!t]
	\centering
	\includegraphics[width=2.7in]{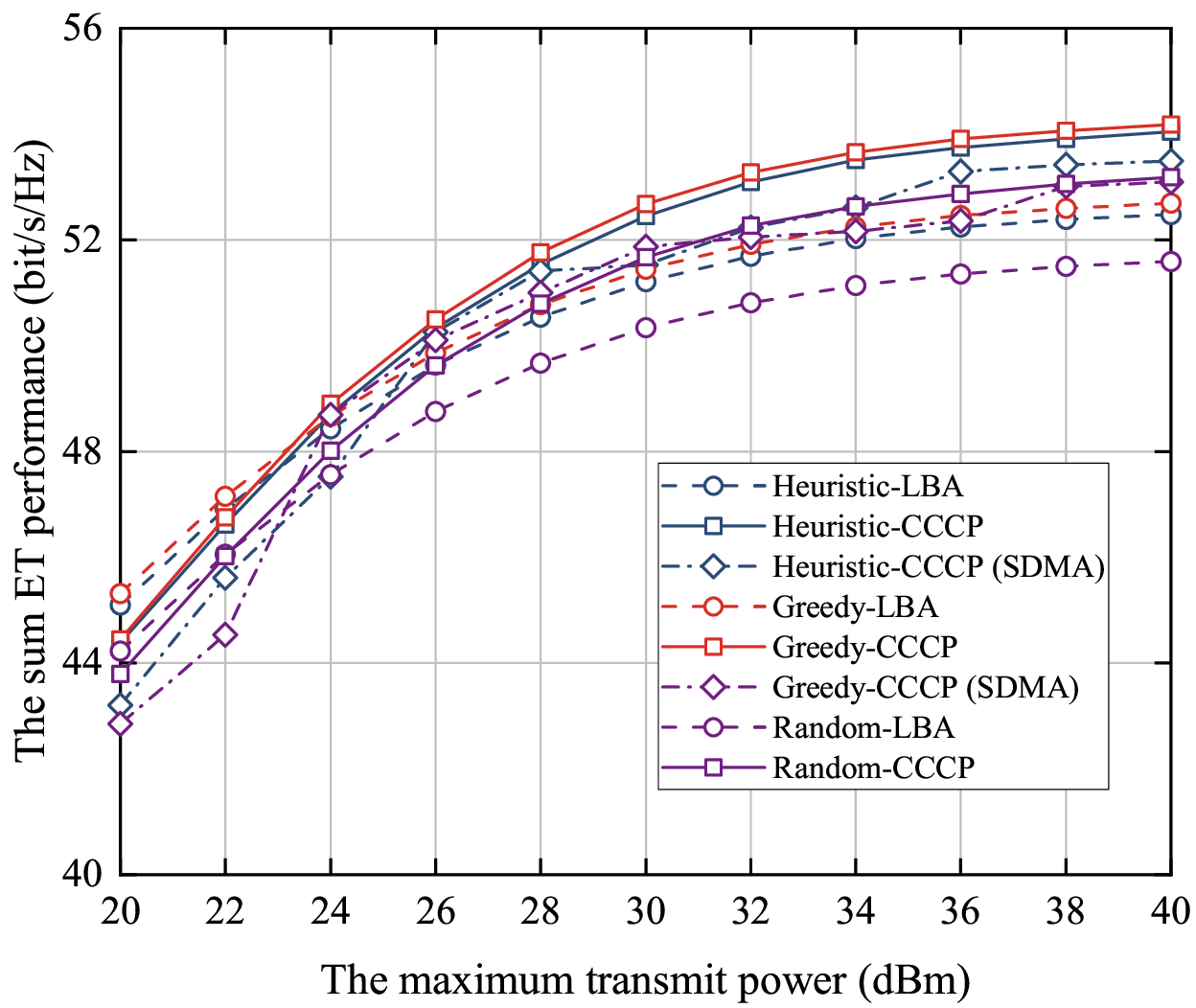}
	\caption{The sum ET versus the maximum transmit power $P_{\max}$.}
	\label{PowerG}
\end{figure}

\begin{figure}[!t]
	\centering
	\includegraphics[width=2.7in]{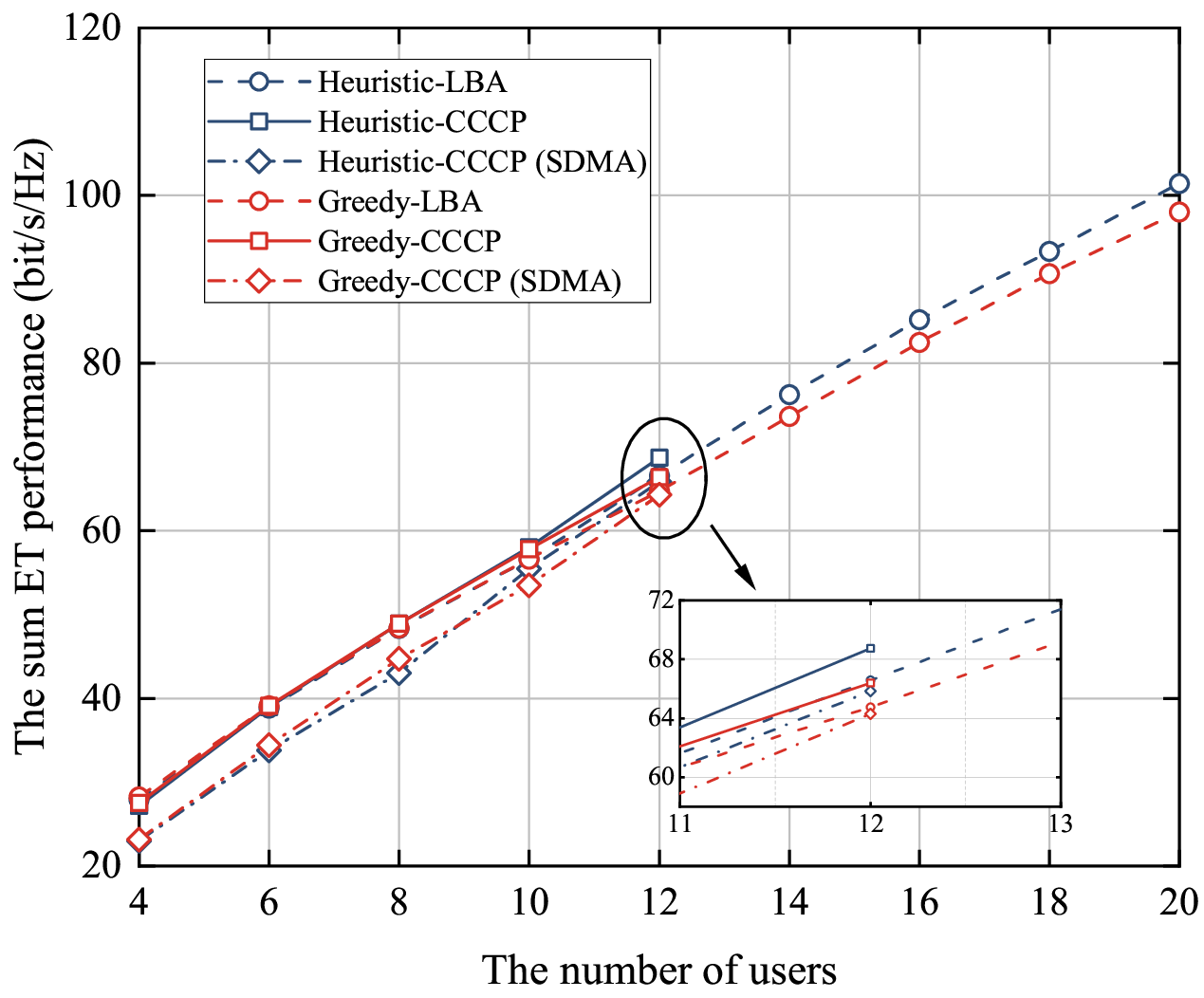}
	\caption{The sum ET versus the number of users $K$.}
	\label{UserG}
\end{figure}

To obtain insights into the impacts of maximum transmit power on ET performance, Fig. \ref{PowerG} presents the relation between ET and $P_{\max}$, where the performance of three different user grouping schemes with LBA or CCCP is evaluated. It has been clearly shown that ET increases as the $P_{\max}$ increases. The greedy search with the CCCP method achieves the highest performance, and then the heuristic grouping with the CCCP method. The performance gap between these two user grouping schemes is small when compared to that of the random method, which demonstrates the high efficiency of the greedy grouping and heuristic grouping methods in user grouping. In addition, we observe that the CCCP method achieves better performance when compared to the LBA method under the same user grouping scheme. For example, when $P_{\max}=30$ dBm, the sum ET of the CCCP method is 2.41\%, 2.40\%, and 2.64\% higher than that of the LBA method for heuristic grouping, greedy grouping, and random methods, respectively. However, considering that there are no outer-layer iterations required for the LBA method, it is worthwhile to deploy this method to avoid additional processing delays. Apart from these, the advantages of RSMA are also verified via the data presented in Fig. \ref{PowerG} since the ETs achieved by SDMA are lower than those achieved by RSMA under the same algorithm configurations.

Fig. \ref{UserG} presents the ET performance versus the number of users $K$. To reduce processing time, we only employ the CCCP method when $K \le 12$. As expected, the ETs achieved by all schemes increase as the number of users. This is because when $M_t \ge K$, the multi-user gain increases with $K$, which brings a higher performance gain to the system. When compared the performance between the heuristic grouping method and greedy grouping method, we find that when $K \le 8$, the ETs achieved by these two methods are very similar. However, when $K$ increases, the heuristic grouping method performs better than the greedy method. This observation verifies the effectiveness of the proposed heuristic method in user grouping. In addition, we observe that the performance gap between the CCCP method and the LBA method is extremely small when $K \le 8$. But this gap also increases with $K$. Furthermore, the superiority of RSMA is also illustrated in this figure, especially when $K$ is small.

\begin{figure}[!t]
	\centering
	\includegraphics[width=2.7in]{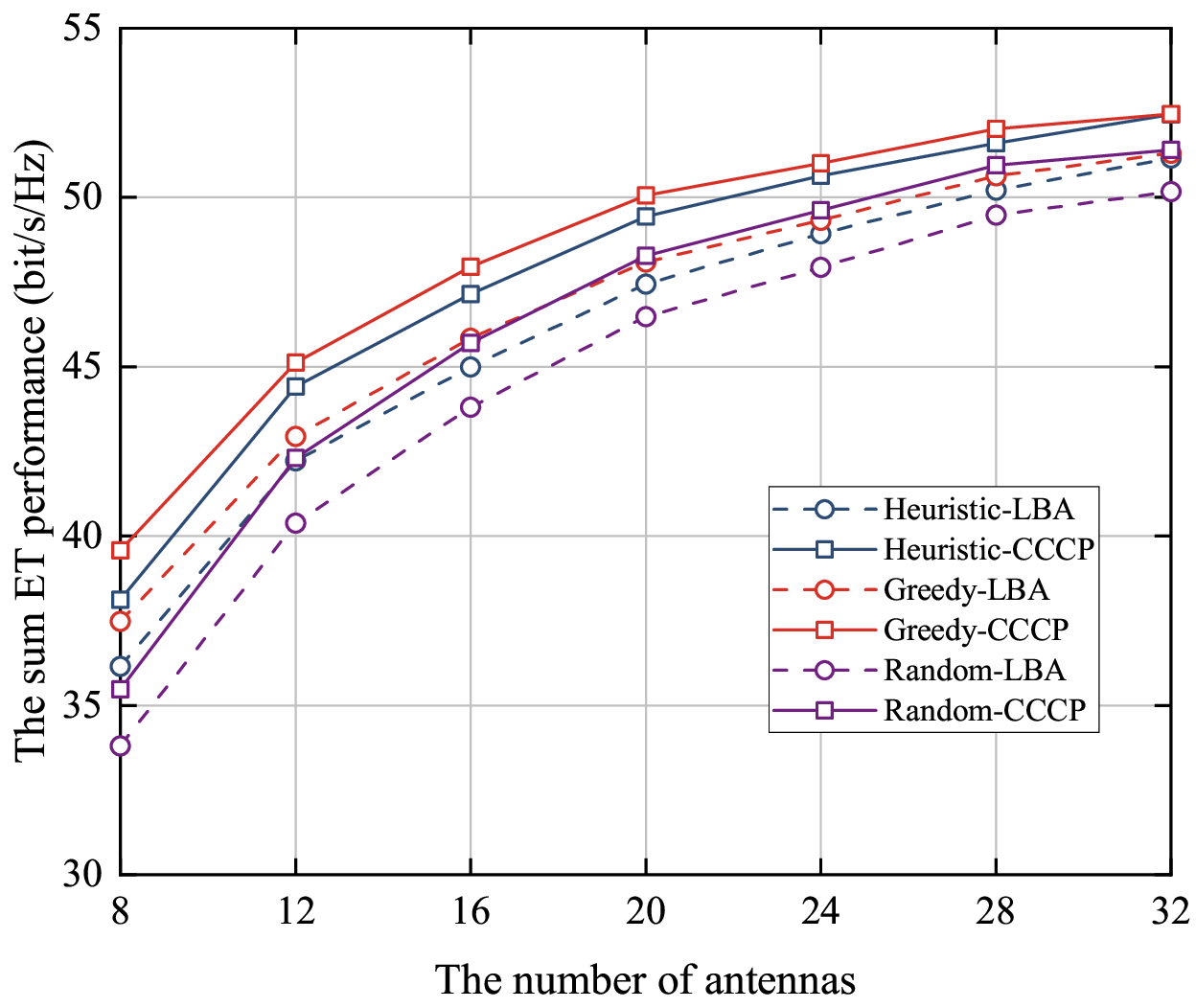}
	\caption{The sum ET versus the number of antennas $M_t$.}
	\label{MtG}
\end{figure}

\begin{figure}[!t]
	\centering
	\includegraphics[width=2.7in]{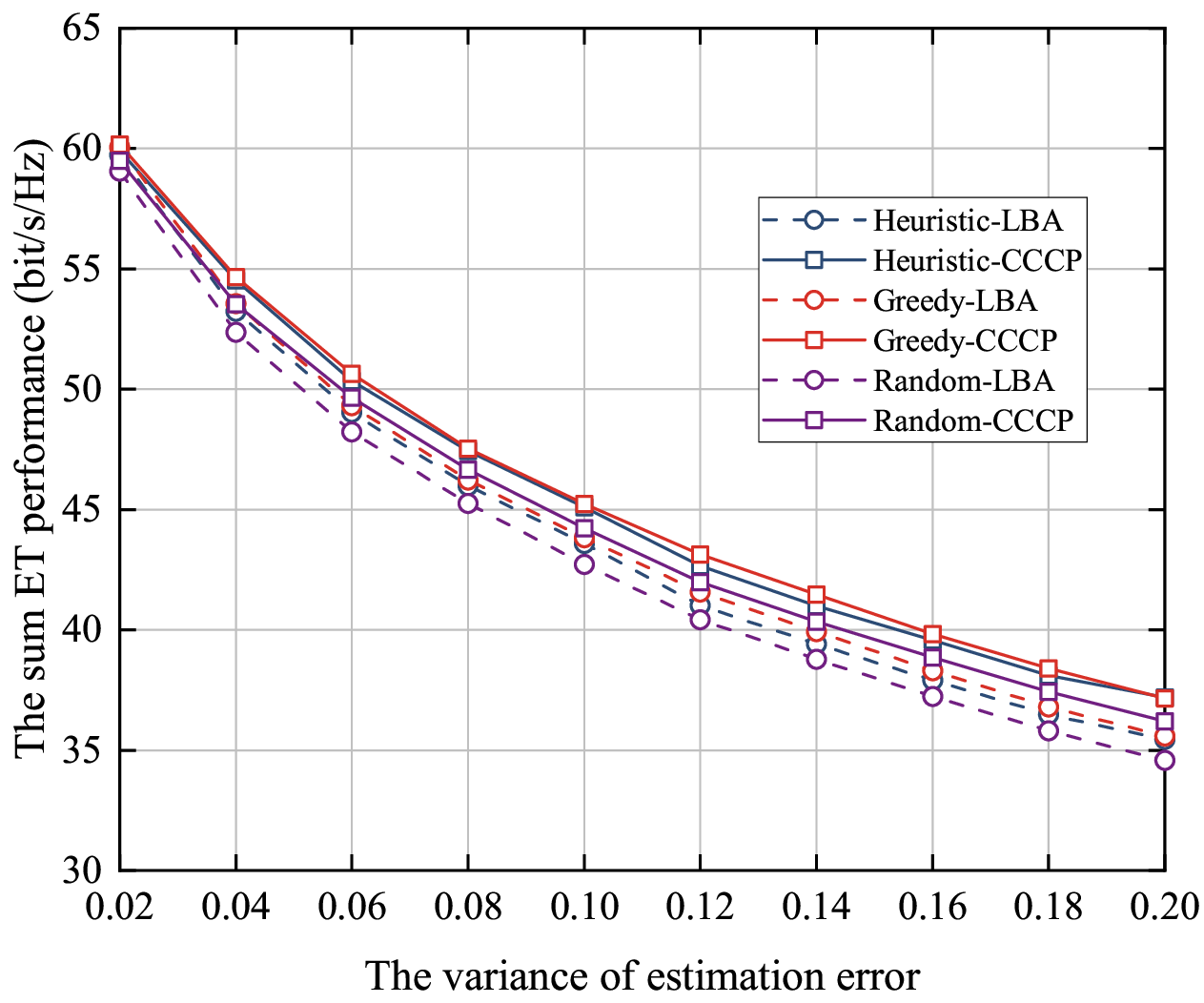}
	\caption{The sum ET of all subcarriers versus the variance of the estimation error $\sigma_{e}^2$.}
	\label{EG}
\end{figure}

Then, the ET performance versus the number of antennas $M_t$ is provided in Fig. \ref{MtG}. In line with our speculation, the ETs achieved by all schemes increase as $M_t$ increases from 8 to 32, and this increasing trend is gradually stabilized. As is shown in this figure, although the performance of the greedy grouping method is higher than that of the heuristic grouping method, this performance advantage gradually decreases with the number of antennas. The performance of the random method is much lower than that of both two other user grouping methods. In addition, we observe that the ET obtained via the LBA method is lower than that obtained through the CCCP method. However, this performance gap decreases with the increasing $M_t$. For the greedy grouping method, when $M_t=8$ and $M_t = 32$, the performance of the CCCP method is 5.44\% and 1.59\% higher than that of the LBA method, respectively. Therefore, the LBA method is more efficient when the number of antennas is large.

\begin{figure}[!t]
	\centering
	\includegraphics[width=2.72in]{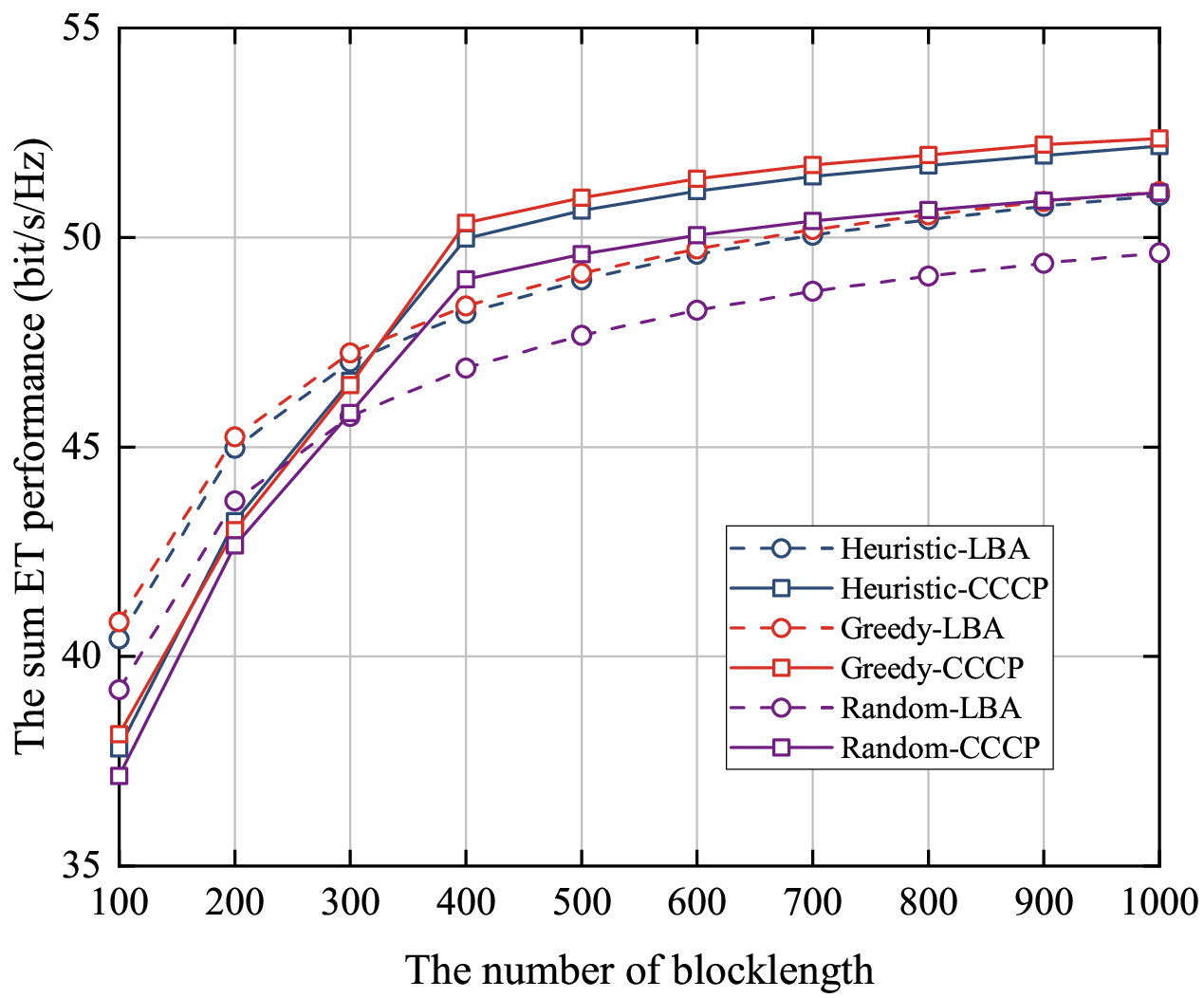}
	\caption{The sum ET versus the number of blocklength $N^{\mathrm{th}}$.}
	\label{BlockG}
\end{figure}

To investigate the impact of the channel uncertainty on system performance, we plot curves of sum ET versus $\sigma_e^2$ in Fig. \ref{EG}. It is evident that the higher $\sigma_e^2$ corresponds to higher uncertainty and leads to worse performance. Similarly, for three different user grouping schemes, the greedy grouping achieves the best performance, and the random method achieves the worst performance. Meanwhile, the CCCP method performs better than the LBA method, and this performance advantage increases with $\sigma_e^2$. For example, when $\sigma_e^2 = 0.02$, the ETs achieved by the CCCP method are  0.31\%, 0.19\%, and 0.76\% higher than that achieved by the LBA method for heuristic grouping, greedy grouping, and random method, respectively. But when $\sigma_e^2 = 0.2$, these percentages increase to 4.89\%, 4.42\%, and 4.67\%, respectively. This finding indicates that the LBA method may not be suitable for systems with higher degrees of channel uncertainty.

Next, Fig. \ref{BlockG} plots the sum ET versus the number of blocklength $N^{\mathrm{th}}$. For a fixed bandwidth $B=1$ MHz, the $N^{\mathrm{th}}$ grows from $100$ to $1,000$, corresponding to $D^{\mathrm{th}}$ increasing from 0.1 to 1 ms. The ETs obtained by all schemes increase as the available blocklength increases, which is in line with the relation given in (\ref{eq1}). The greedy grouping and heuristic grouping methods achieve better performance compared to the random method. However, in Fig. \ref{BlockG}, the ETs obtained via the CCCP method are not always higher than those obtained by the LBA method. The performance advantages of the CCCP method are effective when $N^{\mathrm{th}}\ge 400$. This is because the optimum obtained by the iterative CCCP method is highly dependent on the initial value. Therefore, the LBA method could be more efficient under this circumstance.

\begin{figure}[!t]
	\centering
	\includegraphics[width=2.7in]{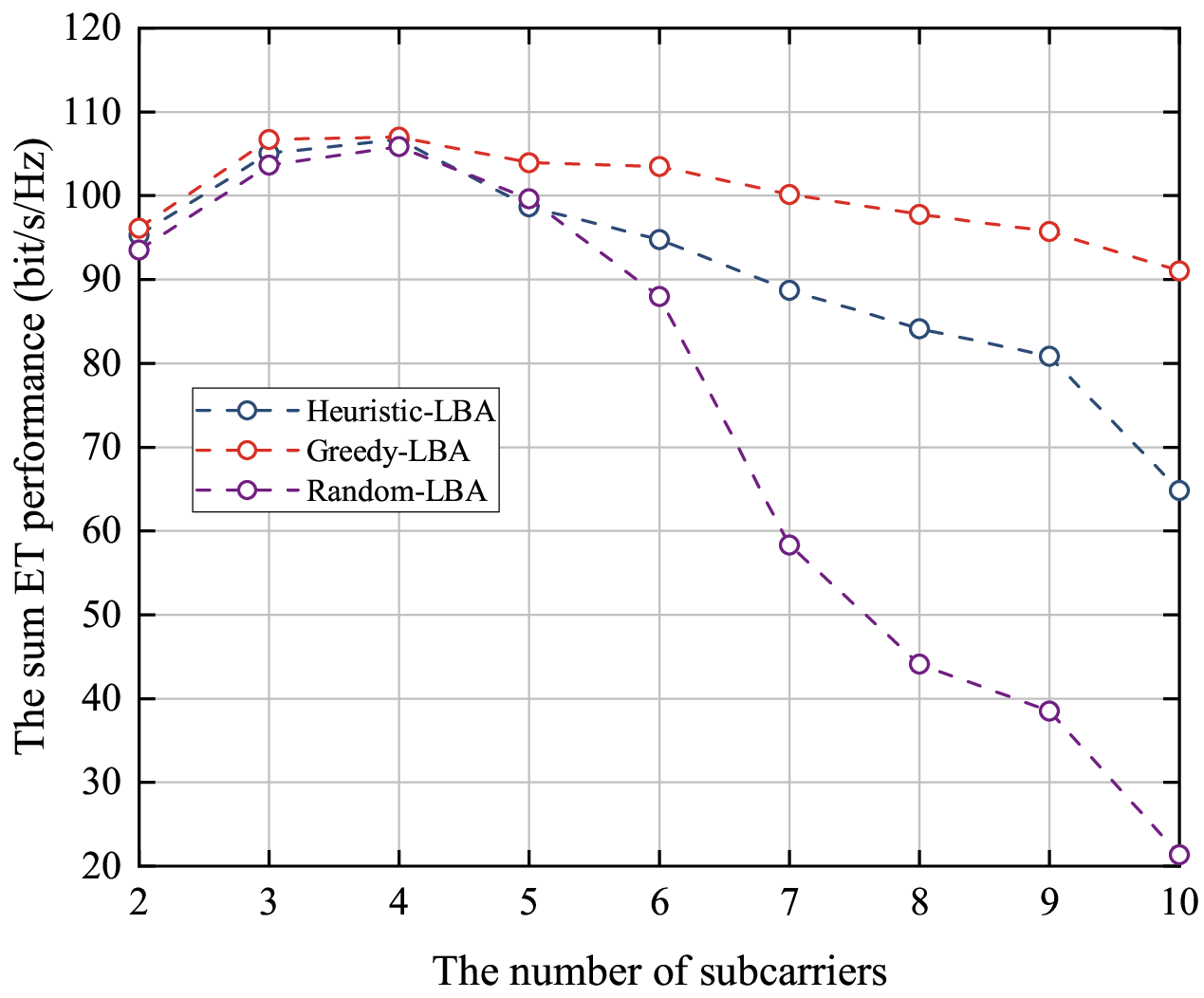}
	\caption{The sum ET versus the number of subcarriers $J$, given $K=20$.}
	\label{CarrierG}
\end{figure}

Fig.~\ref{CarrierG} is provided to evaluate the relation between the sum ET and the number of subcarriers $J$ under a fixed $K=20$. It shows that the sum ET initially increases with increasing $J$ and then decreases. As illustrated in Fig.~\ref{UserG}, the number of multiplexed users decreases with the increasing $J$, which degrades the multi-user gain within a single subcarrier. However, when $J$ is small, e.g., $J=2$, the increase in interference within the subcarrier due to a larger number of multiplexed users reduces the sum ET. Meanwhile, to ensure the effectiveness of ZFBF, the number of antennas must exceed the number of users within a single subcarrier. Therefore, an increase in the number of users within a single subcarrier also requires an increase in the number of antennas, which imposes additional requirements on the system. Taking into account these factors, the trade-off between the number of subcarriers and the number of users deserves to be investigated in MC systems.

\section{Conclusions}\label{Section VI}
This paper investigated the joint user scheduling, power allocation, and rate control for an ET maximization problem in MC-RSMA URLLC systems with imperfect CSIT and SIC. To solve this problem, we developed a decomposition method, where the user scheduling problem and the joint power allocation and rate control problem were separately tackled. Specifically, we first proposed the CCCP and LBA methods for joint power allocation and rate control under a given user grouping scheme. Then, based on the optimized results, a greedy grouping method and a heuristic grouping method were proposed to solve the user grouping problem. The simulation results presented in this paper verified the effectiveness of the CCCP and LBA methods in power allocation and rate control. Meanwhile, the performance superiority of the heuristic grouping and greedy grouping methods was demonstrated when compared to the benchmark. In addition, the performance superiority of RSMA in URLLC services was also demonstrated compared to SDMA. This work proposes a solution for implementing green and sustainable RSMA in URLLC. However, a part of practical communication factors, e.g., the impact of frequency and hardware impairment, can be further investigated in future research.

{\appendix[Proof of Lemma 1]
We first prove the tightness of approximation $\overline{T}_{j,k,c}=R_{j,k,c}(1-\varepsilon^{\mathrm{th}})$ and then prove the tightness of approximation $\overline{T}_{j,k,p}$ using the similar method. To begin with, we assume that $k'=\arg\min_{k\in \mathcal{I}_j}\{R_{j,k,c}^{\mathrm{th}}\}$. Then, from constraint (\ref{14b}), we have
\begin{equation}\label{a1}
\sum_{k\in\mathcal{K}}R_{j,k,c} \le R_{j,c} \le R_{j,k',c}^{\mathrm{th}}.
\end{equation}
Without loss of generality, by letting $R_{j,k',c} = R_{j,c}$ and $R_{j,k,c} = 0\,(k\in \mathcal{I}_j \setminus k')$, $T_{j,k',c} = R_{j,c}(1 - \varepsilon_{j,k',c}(\gamma_{j,k',c},R_{j,c}))$ is obtained.

Next, we prove the monotonicity of $T_{j,k',c}$ w.r.t. $R_{j,c}$. Specifically, the first-order derivative of $T_{j,k',c}$ w.r.t. $R_{j,c}$ is obtained as follows:
\begin{align}\label{a2}
\frac{\partial T_{j,k',c}}{\partial R_{j,c}} 
&= 1 - \varepsilon_{j,k',c} - \frac{\ln2\sqrt{N_j}R_{j,c}}{\sqrt{2\pi V_{j,k',c}}}e^{\frac{-f_{j,k',c}^2}{2}} \notag \\
& \overset{(a)}{\ge} C_1 \!-\! \left(\frac{\ln2\sqrt{N_j}R_{j,c}}{\sqrt{2\pi V_{j,k',c}}}\right)e^{\frac{-f_{j,k,p}^2}{2}} \notag \\
& \overset{(b)}{\ge} C_1 \!- \!\left(\frac{\ln2\sqrt{N_j}g_2(\gamma_{j,k',c})}{\sqrt{2\pi}} \!-\! \frac{Q^{-1}(\varepsilon^{\mathrm{th}})}{\sqrt{2\pi}}\right) \!e^{\frac{-f_{j,k',c}^2}{2}} \notag \\
& \overset{(c)}\ge C_1 \!- \!C_2\left(\frac{\ln2\sqrt{N_j}g_2(\gamma_{j,k',c})}{\sqrt{2\pi}} \!-\! \frac{Q^{-1}(\varepsilon^{\mathrm{th}})}{\sqrt{2\pi}}\right) \notag \\
& \overset{\triangle}{=} g_1(\gamma_{j,k',c}),
\end{align}
where (a) holds since $C_1 = 1 - \varepsilon^{\mathrm{th}} \le 1 - \varepsilon_{j,k'S,c}$; (b) holds since $R_{j,k',c}\le R_{j,k',c}^{\mathrm{th}} = \log_2(1+\gamma_{j,k',c}) - \sqrt{V_{j,k',c}}Q^{-1}(\varepsilon^{\mathrm{th}})/(\sqrt{N_j}\ln(2))$; (c) holds since $C_2 = e^{-\frac{[Q^{-1}(\varepsilon^{th})]^2}{2}}\ge e^{\frac{-f_{j,k',c}^2}{2}}$. Besides, $g_2(\gamma_{j,k',c})=\log_2(1+\gamma_{j,k',c})/\sqrt{V_{j,k',c}}$. Accordingly, the first-order derivative of $g_2(\gamma_{j,k',c})$ is obtained as
\begin{equation}\label{a3}
g_2'(\gamma_{j,k',c}) = \frac{\gamma_{j,k',c}(\gamma_{j,k',c}+2) - \ln(1+\gamma_{j,k',c})}{\ln(2)V_{j,k',c}^{3/2}(1+\gamma_{j,k',c})^3} \ge 0.
\end{equation}
Hence, $g_2(\gamma_{j,k',c})$ is an increasing function of $\gamma_{j,k',c}$, which means that $g_1(\gamma_{j,k',c})$ decreases with increasing $\gamma_{j,k',c}$.

Next, letting $g_1(\gamma_{j,k',c}) \le 0$, $g_2(\gamma_{j,k',c})$ should thus satisfy
\begin{equation}\label{a4}
g_2(\gamma_{j,k',c}) \ge \frac{\frac{2\pi C_1}{C_2} + Q^{-1}(\varepsilon^{th})}{\ln2\sqrt{N_j}}.
\end{equation}
It is evident that for a given $N_j$ and $\varepsilon^{\mathrm{th}}$, the right-hand side of the above inequality is a constant. Therefore, considering the QoS constraints for a typical URLLC scenario, i.e., $\varepsilon^{\mathrm{th}} \le 10^{-4}$ and $N_j \le 10^4$, we have $g_2(\gamma_{j,k',c}) \ge  91$. Since $g_2(\gamma_{j,k',c})$ increases with $\gamma_{j,k',c}$, when $\gamma_{j,k',c}$ goes large, we have $V_{j,k',c}\approx 1$. Thus, $\gamma_{j,k',c}$ satisfies $\gamma_{j,k',c} = \gamma_{j,k',c}^{\dag}\ge 2^{91}- 1$. However, this SINR requirement is impractical in URLLC under imperfect CSIT and SIC. Considering this, $g_1(\gamma_{j,k,c}) > 0$ can be guaranteed when $\varepsilon^{\mathrm{th}} \le 10^{-4}$ and $N_j \le 10^4$. As a result, $T_{j,k',c}$ is a monotone increasing function of $R_{j,c}$.

Therefore, $R_{j,c} = R_{j,k',c}^{\mathrm{th}}$ always holds when $\gamma_{j,k',c}\le 2^{91}-1$, which means that
\begin{equation}\label{a5}
\max_{R_{j,c}}\left\{T_{j,k',c} \right\}= \max_{R_{j,k,c}}\left\{\overline{T}_{j,k',c }\right\}= R_{j,k',c}^{\mathrm{th}}(1 - \varepsilon^{\mathrm{th}}).
\end{equation}
Note that $\varepsilon_{j,k,c}(\gamma_{j,k,c}, R_{j,k',c}^{\mathrm{th}})\le\varepsilon^{\mathrm{th}}$ since $R_{j,k,c}^{\mathrm{th}}\ge R_{j,k',c}^{\mathrm{th}}$. For $k\in \mathcal{I}_j \setminus k'$, we have
\begin{align}\label{a6}
\max_{R_{j,k,c}}\left\{T_{j,k,c} \right\}- \max_{R_{j,k,c}}\left\{\overline{T}_{j,k,c }\right\}&= R_{j,k,c}^{\mathrm{th}}(\varepsilon^{th} - \varepsilon_{j,k,c}(R_{j,k',c}^{\mathrm{th}}))
\notag \\
&\le R_{j,k,c}^{\mathrm{th}}\varepsilon^{\mathrm{th}}.
\end{align}
Due to a extreme small $\varepsilon^{\mathrm{th}}$, we conclude the approximation $\overline{T}_{j,k,c}$ is tight at the optimal point $R_{j,k,c}^{\mathrm{th}}$. Therefore, $T_{j,k,p}$ can be expressed as $T_{j,k,p} = R_{j,k,p}(1 - \varepsilon_{j,k,p}-\varepsilon^{\mathrm{th}})$. Based on this, the tightness of approximation $\overline{R}_{j,k,p}$ can be demonstrated by using similar steps (\ref{a2})-(\ref{a6}), i.e., we have
\begin{equation}\label{a7}
	\max_{R_{j,k,p}}\left\{T_{j,k,p} \right\}= \max_{R_{j,k,p}}\left\{\overline{T}_{j,k,p }\right\}= R_{j,k,p}^{\mathrm{th}}(1 - 2\varepsilon^{\mathrm{th}}),
\end{equation}
which completes the proof of Lemma 1.

}

\newpage



\vfill

\end{document}